\documentclass[showkeys,preprintnumbers,amsmath,amssymb,nofootinbib]{revtex4}

\usepackage{graphicx}
\usepackage{dcolumn}
\usepackage{bm}
\usepackage{amsmath,amssymb,epsfig} 
\usepackage{wrapfig}
\usepackage{comment}
\usepackage{paralist}

\begin{document}

\title{Corner-cube retro-reflector instrument for advanced lunar laser ranging}

\author{Slava G. Turyshev, James G. Williams, William M. Folkner, \\ Gary M. Gutt, Richard T. Baran, Randall C. Hein, Ruwan P. Somawardhana 
}

\affiliation{\vskip 3pt
Jet Propulsion Laboratory, California Institute of Technology,\\
4800 Oak Grove Drive, Pasadena, CA 91109-0899, USA
}

\author{John A. Lipa and Suwen Wang
}

\affiliation{%
Hansen Experimental Physics Laboratory, Department of Physics,
Stanford University, Stanford, CA 94305-4085, USA
}%

\date{\today}

\begin{abstract}
Lunar laser ranging (LLR) has made major contributions to our understanding of the Moon's internal structure and the dynamics of the Earth-Moon system. Because of the recent improvements of the ground-based laser ranging facilities, the present LLR measurement accuracy is limited by the retro-reflectors currently on the lunar surface, which are arrays of small corner-cubes. Because of lunar librations, the surfaces of these arrays do not, in general, point directly at the Earth. This effect results in a spread of arrival times, because each cube that comprises the retroreflector is at a slightly different distance from the Earth, leading to the reduced ranging accuracy. Thus, a single, wide aperture corner-cube could have a clear advantage. In addition, after nearly four decades of successful operations the retro-reflectors arrays currently on the Moon started to show performance degradation; as a result, they yield still useful, but much weaker return signals. Thus, fresh and bright instruments on the lunar surface are needed to continue precision LLR measurements. We have developed a new retro-reflector design to enable advanced LLR operations. It is based on a single, hollow corner cube with a large aperture for which preliminary thermal, mechanical, and optical design and analysis have been performed. The new instrument will be able to reach an Earth-Moon range precision of 1-mm in a single pulse while being subjected to significant thermal variations present on the lunar surface, and will have low mass to allow robotic deployment. Here we report on our design results and instrument development effort.

\end{abstract}
 
\maketitle


\section{\label{sec:intro}Introduction}

\begin{figure}[!h]
    \begin{center} 
 \epsfig{figure=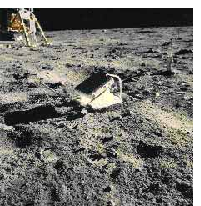,width=43mm} 
\hskip 10pt
\epsfig{file=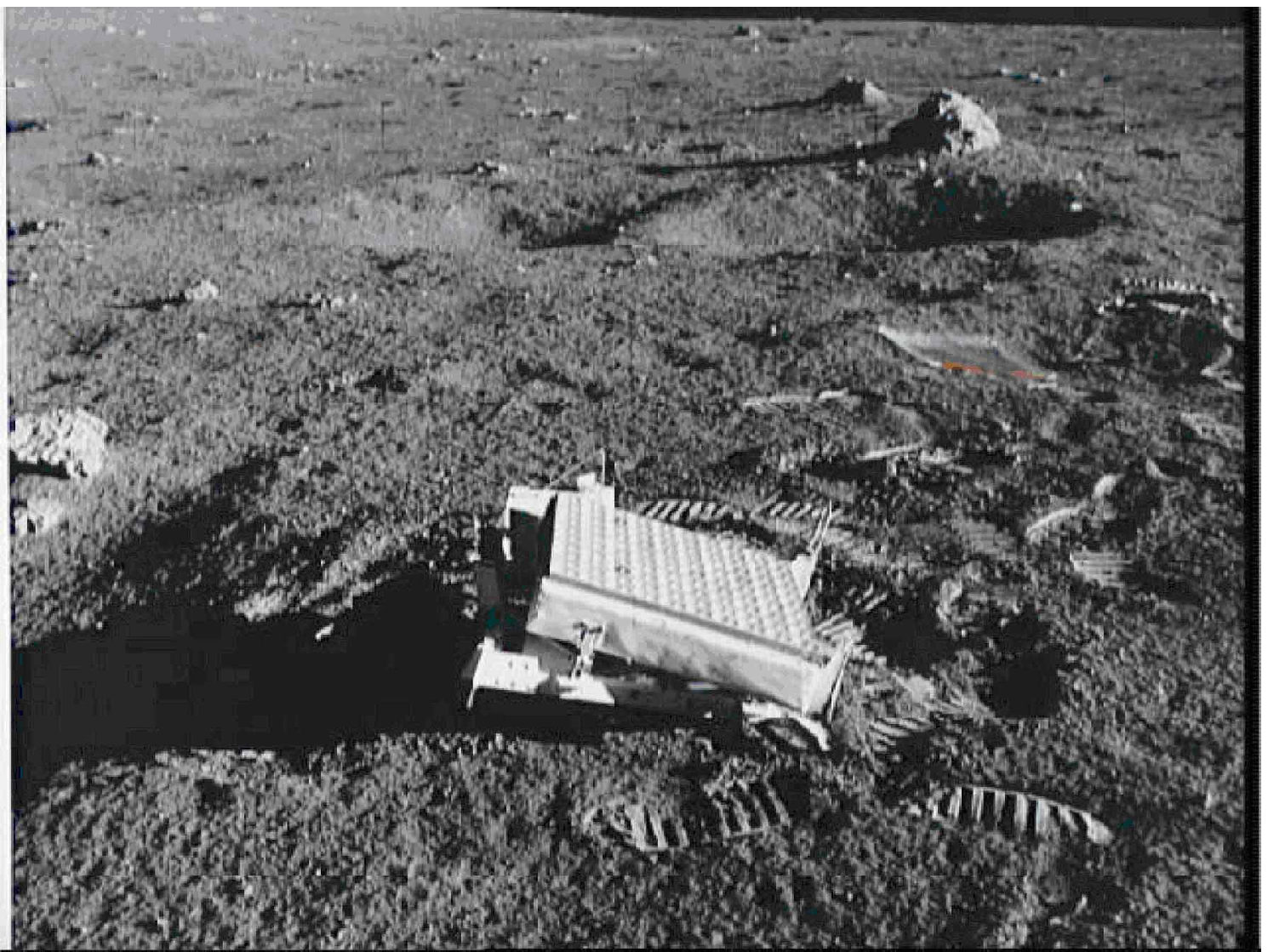,width=59mm}
\caption{Apollo 11 (left) and Apollo 14 (right) laser retro reflector arrays.} 
\label{fig:2}
    \end{center}
  \begin{center}
    \includegraphics[width=0.40\textwidth]{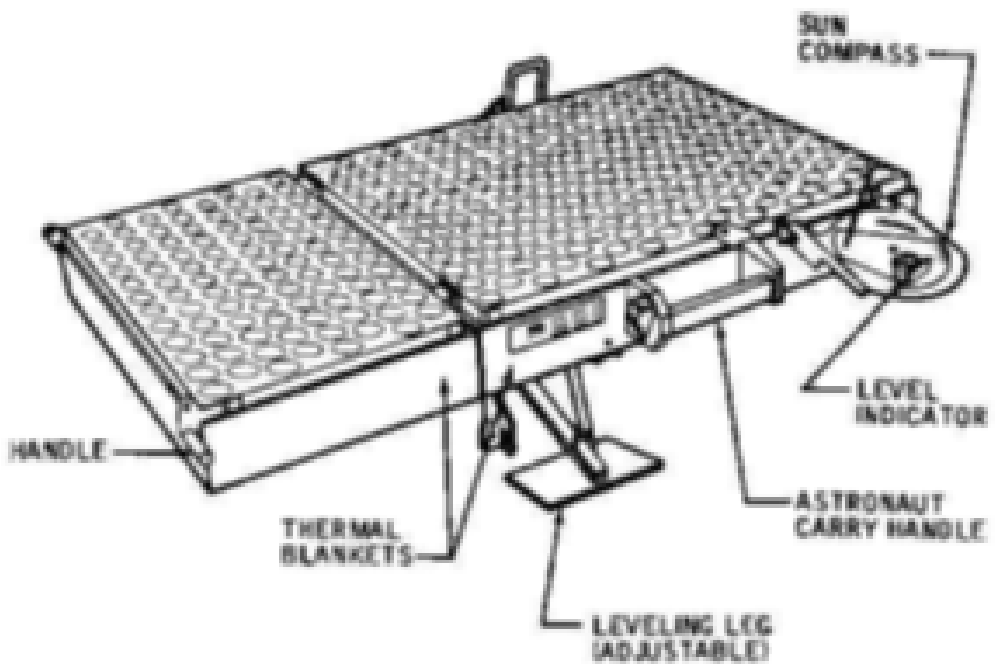}
\hskip 10pt
    \includegraphics[width=0.25\textwidth]{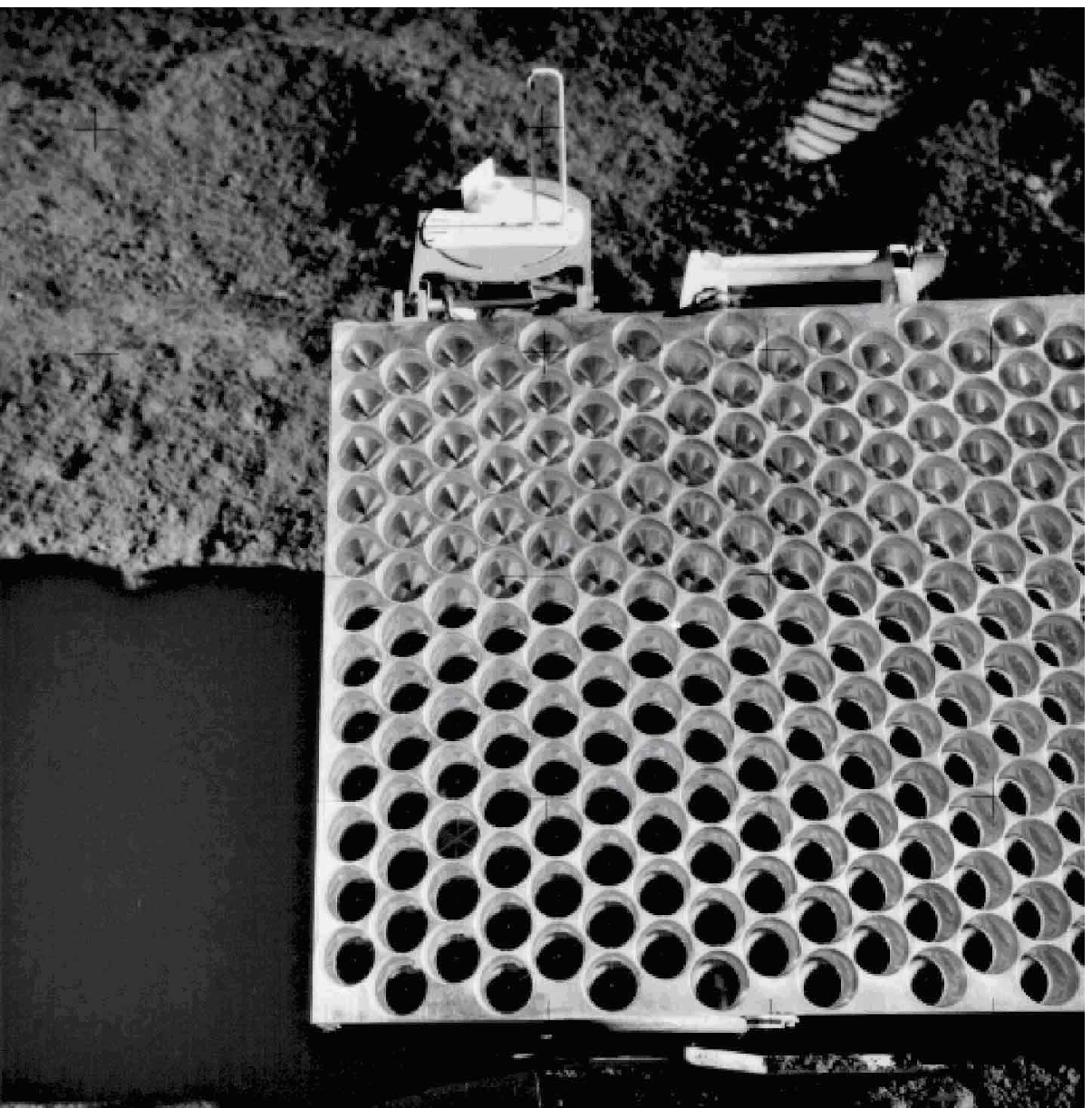}
  \vspace{-5pt}
  \caption{Left: Line drawing of A15 reflector. Note the gap between the two sections. Right: Photo of Apollo 15 retro-reflector array on the lunar surface.}
\label{fig:4}
  \end{center}
%
    \begin{center} 
 \epsfig{figure=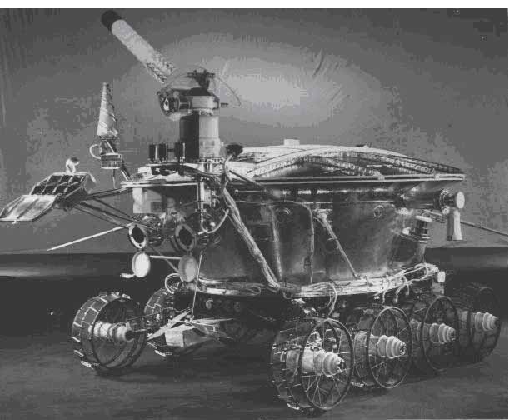,width=58mm} 
\hskip 20pt
 \epsfig{file=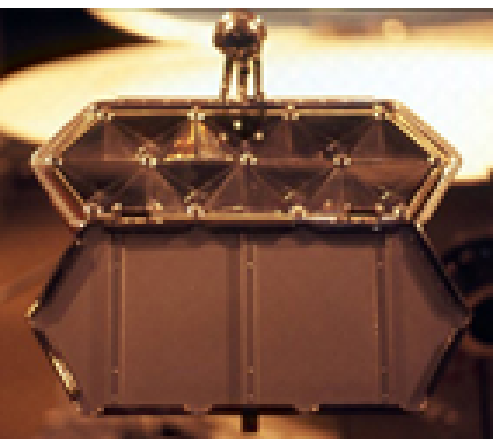,width=54mm}
\caption{Lunokhod 1 with the retroreflector array protruding at far left and a photo of its reflector.} \label{fig:3}
    \end{center}
\end{figure}

\begin{figure}[!t]
    \begin{center} 
\begin{minipage}[t]{.50\linewidth}
 \epsfig{file=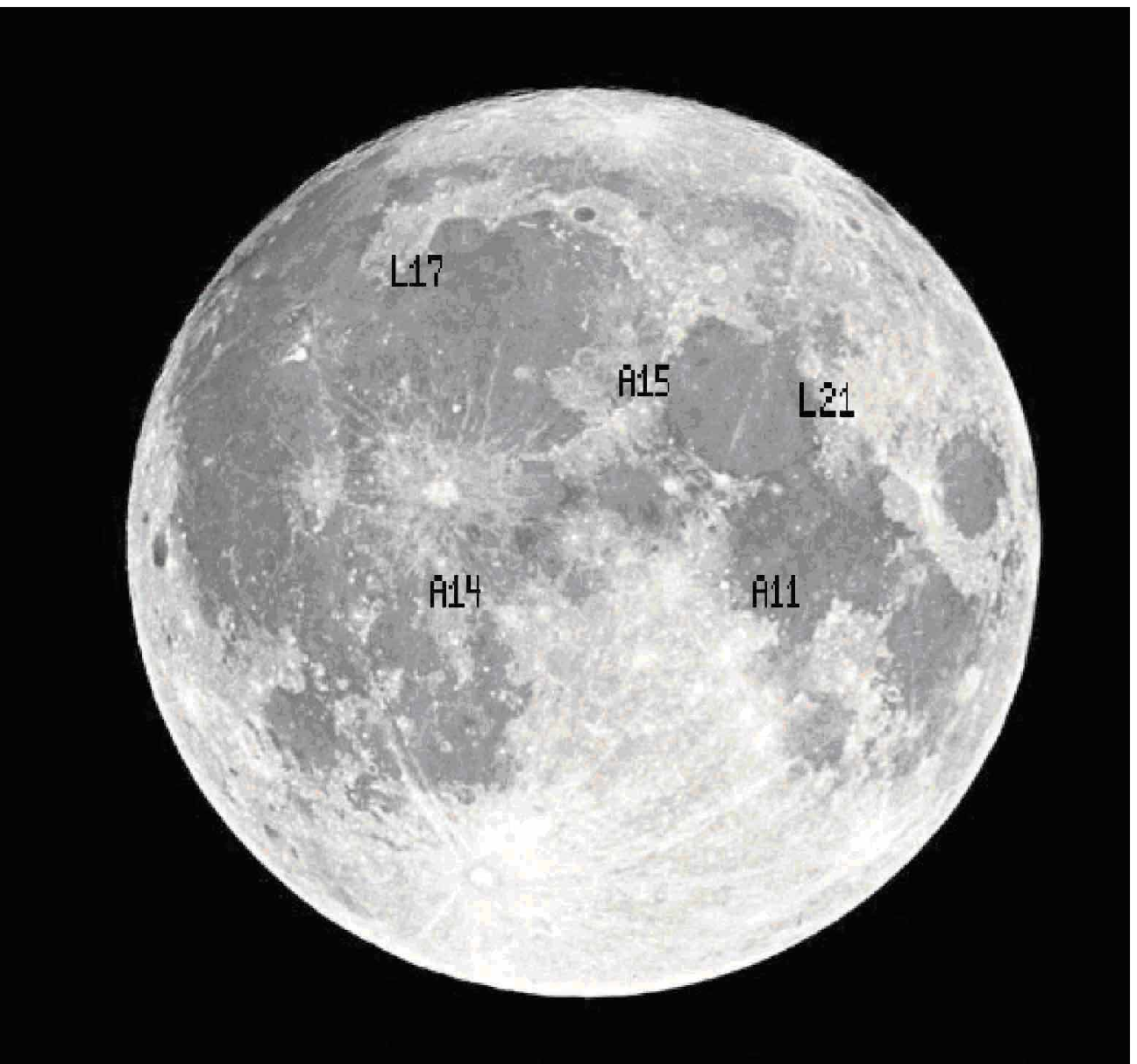,width=53mm} 
\caption{The LLR retro-reflector sites on the Moon.} 
\label{fig:LLR-sites}
    \end{minipage}
\hskip -10pt
\begin{minipage}[t]{.42\linewidth} 
\epsfig{file=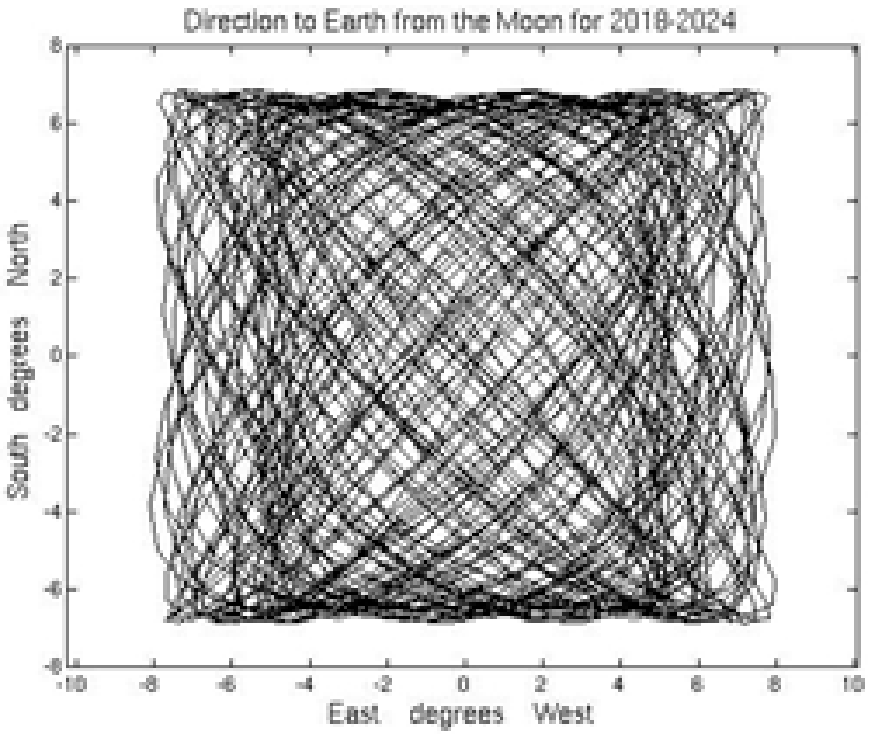,width=60mm}
\caption{Variation of Earth center about mean direction for 6 years, 2018 to 2024, as seen from Moon.} \label{fig:1-librations}
    \end{minipage}
    \end{center}
  \vspace{-20pt}
\end{figure}

Since its initiation in 1969, lunar laser ranging (LLR) has contributed significantly to our understanding of the Moon's internal structure and the dynamics of the Earth-Moon system. After initial deployment of the Apollo 11 corner cube retro-reflector (CCR) array \cite{Bender-etal-1973,Williams-etal-2009} two more CCR packages were set up by Apollo 14 and 15 (Figs.~\ref{fig:2}-\ref{fig:4}). In addition, two French-built CCR arrays were on the Lunokhod 1 and 2 rovers placed on the Moon by the Soviet Luna 17 and 21 missions, respectively (Fig.~\ref{fig:3}). Figure~\ref{fig:LLR-sites} shows the five LLR reflector sites on the Moon \cite{Faller-etal-1971,Faller-1972,Faller-etal-1972,Fournet-1972}. 

LLR data provide insight over broad areas spanning lunar science, Earth sciences, geodesy and geodynamics, solar system ephemerides, gravitational physics, and terrestrial and celestial reference frames \cite{Williams-Newhall-Dickey-1996,Williams-Newhall-Dickey-1996-2}. LLR retro-reflectors at a variety of locations on the Moon allow measurements of physical librations, the variations of rotation and pole orientation, and solid-body tides, which provides the science information for reserach on lunar variable gravity and lunar interior with LLR \cite{Williams-etal-2001,Rambaux-Williams-2011}. Clearly, retro-reflectors more widely distributed than the Apollo sites which span about 1/3 the lunar diameter  would improve the sensitivity to physical librations and tides \cite{Turyshev-etal-2008}; however, there is an even greater need for retro-reflectors with much improved ranging precision. This paper is motivated by these important issues. 

The LLR experiment relies on measurements of the round-trip travel time for a laser pulse between transmission of a pulse from an Earth tracking station to a retro-reflector on the lunar surface and back to the tracking station \cite{Williams-etal-2009}. Three different reflector designs were used for the current lunar installations (see Figs. \ref{fig:2}-\ref{fig:3}). Since the details are crucial to our discussion, the designs are summarized here:
\begin{itemize}
\item {\it The Apollo 11 and Apollo 14 reflectors} are square arrays with 100 circular CCRs, each 38 mm in diameter, arranged in 10 rows and 10 columns.  The array size is 46 cm $\times$ 46 cm, with the CCRs recessed by 19 mm \cite{Alley-etal-1969,Faller-etal-1971}.  Photos of the Apollo 11 and 14 reflectors on the Moon are shown in Fig.~\ref{fig:2}.

\item {\it The Apollo 15 reflector} is a larger array with 300 CCRs in two panels, using hexagonal packing.  Each CCR is similar to those of Apollo 11 and 14, that is 38 mm in diameter and recessed.  One panel has 204 reflectors in 17 columns of 12 rows, then there is a gap, followed by another panel of 96 reflectors in 8 columns of 12 rows.  The total array dimensions are 1.05 m $\times$ 0.64 m \cite{Faller-etal-1972}. On the left part in Fig.~\ref{fig:4} is a drawing of the entire apparatus. On the right 
is a photograph of the reflector on the Moon, showing the hexagonal packing.

\item {\it The Lunokhod reflectors} feature 14 triangular corner cubes, arranged in two rows in hexagonal packing \cite{Fournet-1972}. Each triangle is about 106 mm on a side.\footnote{Note that the hexagonal design yields 6 fold symmetry rather than the 4 fold of the Apollo 11 design.}  The Lunokhod reflectors were aligned to Earth by maneuvering the Soviet-built rovers, thus, yielding alignment less precise compared to that of the Apollo arrays. Since the Lunokhod reflectors are not expected to be as well aligned to Earth (the Lunokhod 2 reflector is likely misaligned by $\sim$5$^\circ$ with respect to the mean Earth direction \cite{Samain-etal-1998}), there may be times when the deviation from normal incidence is higher compared to better aligned Apollo arrays (which likely are within $\sim1^\circ$ with respect to the mean Earth direction \cite{Samain-etal-1998}). A photo of a French-built Lunokhod retroreflector is shown in Fig.~\ref{fig:3}.
\end{itemize}
 
Most LLR scientific results come from long observing campaigns at several observatories. One station is the McDonald Laser Ranging System (MLRS)  in Texas, USA \cite{Shelus-etal-2003}. This 0.76 m diameter station has been ranging from its current site since 1988. Earlier data is available from other McDonald sites back to 1969. Another station is at the Observatoire de la C\^ote d'Azur (OCA)  in France \cite{Samain-etal-1998} which began accurate observations in 1984. Both stations operate in a multiple-target mode, observing artificial satellites in addition to the lunar CCR arrays. The Apache Point Observatory Lunar Laser-ranging Operations (APOLLO)  started ranging in 2006 \cite{Murphy-etal-2007,Murphy-etal-2008}. With a 3.5 m diameter telescope, this facility was designed to achieve mm range precision and to enable corresponding order-of-magnitude gains in the determination of relevant physics parameters \cite{Williams-Turyshev-Murphy-2004}. 

The measurement accuracy was originally limited by timing accuracy and the length of the laser pulse transmitted. Several decades of technology development have since made accurate timing and very short, high-power laser pulses possible.  As a result, even at the beginning of the APOLLO operations \cite{Murphy-etal-2008}, it became clear that the LLR measurement accuracy was limited by the retro-reflectors currently on the lunar surface. These instruments are arrays of small CCRs and, since range measurements are based on single-photon timing events, there is an ambiguity in knowing which CCR the photon was reflected from. This ambiguity results in the increase of the width of the returned pulses, thereby giving rise to the largest component in the ranging error budget. A fraction of this error can be statistically corrected by collecting a large number of photons, but the remaining part is left in the data, limiting the progress in LLR science. Thus, it seems natural that with the upgrade of the ground-based LLR capabilities, the next logical step is to improve the quality and distribution of the lunar CCRs, possibly using the approach described in this paper.

Recently, a design for a 10 cm solid CCR prism was presented \cite{Currie-etal-2009,Currie-etal-2011}. Although this instrument would allow for a range precision below 1 mm, it would have a signal strength of only 0.2 when compared to the fresh small Apollo 11 or 14 arrays and would require an astronaut-assisted deployment. Such a low signal strength is undesirable, and human deployment is unlikely in the near future.

We present here the design and supporting analysis of a new, high-precision, retro-reflector instrument using a single, hollow corner cube with a 170 mm aperture capable of reaching the 1-mm {\it single pulse} Earth-Moon range precision (compared to a few cm currently) needed to support advanced LLR operations.  The instrument will be as bright as fresh Apollo 11 and 14 arrays, is expected to perform under the harsh environment on the lunar surface (e.g., dust, significant thermal variations), will have low mass, and will allow for robotic deployment. This will allow an improvement of the LLR measurement accuracy by more than an order of magnitude, and will have a large enough signal to be easily detected by ground stations. 

This paper is organized as follows: Section \ref{sec:2} discusses the CCR arrays currently on the Moon and addresses their limitations. In Section \ref{sec:3} we describe the design considerations for a new LLR retro-reflector with advanced properties. Section \ref{sec:model-test} discusses modeling and testing of the new instrument. This section also summarizes the current design for the CCR instrument. Section~\ref{sec:development} addresses important considerations relevant to the process of building a CCR instrument with unique performance. We conclude in Section \ref{sec:concl} with a summary and recommendations. 

\section{Existing retro-reflector arrays and their problem} 
\label{sec:2}

The Moon's rotation is locked to the Earth's direction by tidal forces \cite{Williams-Newhall-Dickey-1996-2}. At the same time the Earth is not always in the exact same position, as seen from the lunar surface \cite{Scheffer-2005}. Because the Moon's orbit is eccentric and perturbed, the Earth's center appears to move as much as $8.2^\circ$ east and west around its mean position. Also, due to the tilt of the Moon's orbit with respect to its equator, the Earth's center appears to move about $6.9^\circ$ north and south of its mean position. The combination of these effects is known as lunar librations, which are shown in Figure~\ref{fig:1-librations} for the years 2018-2024. As seen from the Moon, the Earth appears up to $2.0^\circ$ in diameter, so a spot formed by the returning photons on the Earth's surface can appear to be up to $1.0^\circ$ from the center. 

Lunar librations have practical implications for LLR since the reflectors do not track the Earth. For practical purposes, one needs to get enough effective aperture which is achieved by assembling the retro-reflectors in a form of planar arrays of small CCRs. The normal to this plane does not always pass through the ranging telescope, resulting in a spread of the returned pulse, because individual CCRs forming the array are at different distances from the telescope. In other words, when the laser pulse from the Earth has a non-zero angle of incidence on the array (due to lunar librations), the different propagation distances to the vertices of multiple CCRs forming the array result in an increase in the pulse width.  This contributes to a large single-pulse error \cite{Murphy-etal-2008}.

If the reflectors on the Moon were comprised of a single corner cube, this misalignment would matter very little, since corner reflectors by design have equal path lengths to any point on their surface, for any angle. Also, a larger CCR will have a smaller diffraction-limited beam size, giving a higher intensity on return to the Earth.  Thus, even with a smaller optical aperture than the Apollo 11 and 14 retro-reflectors, the signal return of a single, large CCR can equal the return intensity. Below we analyze this problem using historical LLR data. 

\subsection{\label{sec:2:33}Effect of lunar librations on the accuracy of LLR ranges} 

\begin{wraptable}{R}{0.45\textwidth}
\vskip-20pt 
\caption{Retro-reflector returned pulse spread in ps (rms) as a function of the east-west and north-south libration angles, computed for the Apollo 15 array.
\label{tab:1-pulse-spread}}
\begin{ruledtabular}
\begin{tabular}{|c|ccccc|}
8.0$^\circ$	& 145	& 164	& 204	& 256	& 315\\
6.0$^\circ$	& 112	& 132	& 179	& 237	& 300\\
4.0$^\circ$	& 75	& 102	& 159	& 223	& 290\\
2.0$^\circ$	& 37	& 80	& 146	& 214	& 283\\
0.0$^\circ$	& 0.0	& 70	& 141	& 211	& 281\\\hline
NS/EW	& 0.0$^\circ$	& 2.0$^\circ$	& 4.0$^\circ$	& 6.0$^\circ$	& 8.0$^\circ$\\
\end{tabular}
\end{ruledtabular}
\caption{The minimum number of photons needed for a 1 mm normal point uncertainty, as a function of libration angles. Computed by taking the retro-reflector pulse spread from Table~\ref{tab:1-pulse-spread}, adding 50 ps in quadrature for the system response, and then computing how many photons must be averaged 
to bring the rms value down to 6.6 ps.
\label{tab:2-No-photons}}
\begin{ruledtabular}
\begin{tabular}{|c|ccccc|}
8.0$^\circ$	& 563	& 675	& 1008	& 1560	& 2330\\
6.0$^\circ$	& 343	& 456	& 792	& 1350	& 2129\\
4.0$^\circ$	& 185	& 298	& 637	& 1200	& 1984\\
2.0$^\circ$	& 89	& 203	& 543	& 1108	& 1897\\
0.0$^\circ$	& 57	& 171	& 512	& 1078	& 1869\\\hline
NS/EW	& 0.0$^\circ$	& 2.0$^\circ$	& 4.0$^\circ$	& 6.0$^\circ$	& 8.0$^\circ$\\
\end{tabular}
\end{ruledtabular}
\end{wraptable}

To evaluate the tilted array problem, consider that a 1m-sized reflector with an 8.2$^\circ$ misalignment has path lengths that vary by up to $\pm$7 cm, depending on which corner cube reflects a given photon, or $\pm$475 ps spread in arrival times. Assuming each photon has an equal chance of being reflected by each individual CCR, and using the physical dimensions of the Apollo 15 array, one can estimate the pulse spread of the array as a function of the east-west (EW) and north-south (NS) librations \cite{Scheffer-2005}. Results are shown in Table~\ref{tab:1-pulse-spread}.

A normal point is the combination of ranges from multiple single-pulse measurements. The best normal points today have one to several mm of uncertainty \cite{Williams-etal-2009}. This is a product of two effects. First, there is uncertainty for each photon associated with the retro-reflector arrays spreading the pulse. Second, since the strength of the returned signal is so low there are very few photons arrive at the detector. To compensate for these effects and to reduce the rms variation down to 6.6 ps (needed to achieve 1 mm precision in range) one must average thousands of photons.  

Table~\ref{tab:2-No-photons} shows the number of photons, as a function of EW and NS librations, that must be averaged to reach the rms normal point accuracy at the level of 1 mm. It assumes the Apollo 15 reflector and a 50 ps system response. The Apache Point program \cite{Murphy-etal-2009} collects the large numbers of photons needed for the averaging of Table~\ref{tab:2-No-photons} by using a large 3.5~m telescope and a 4$\times$4 array detector. Other LLR stations have smaller telescopes and single detectors and cannot collect the large numbers of photons needed for 1-mm level normal point uncertainty.

As Murphy et al. \cite{Murphy-etal-2008} report, this pulse spread is the largest component of the ranging error budget, since the other errors have been reduced dramatically since the early days of LLR.  The total of all other uncertainties is only 60-75 ps for the OCA station \cite{Samain-etal-1998}, and is estimated as 52 ps for the station at the Apache Point Observatory (APO) \cite{Murphy-etal-2008}. Although one may think about constructing additional LLR facilities with APOLLO-level ranging capabilities, it has become clear that new laser ranging instruments on the Moon with a much improved target precision are needed.

\subsection{\label{sec:2.2} Using historical data to evaluate the problem}
 
When looking at the historical LLR data, one clearly sees that there are large variations in the detected signal strength due to atmospheric scintillation that strongly affects the amount of data that can be taken. Table~\ref{tab:3-LLR-ranges-station} gives the number of normal points, constructed from multiple photon returns, for the six LLR stations in our analysis that have more than a few days of data. The total number of observations for each retro-reflector array is given for each 
site along with the time span of the measurements.

\begin{table}[h!]
\caption{Number and percentage of lunar laser ranges by ranging station and retro-reflector array from 1970 to 2009. The total number of ranges during this time span is 17,107. 
\label{tab:3-LLR-ranges-station}}
\begin{center}
\begin{tabular}{|c|c|c|c|c|c|c|c|c|c|c|c|}\hline 
Station	& Receiver & Time span,  & \multicolumn{2}{c|}{Apollo 11} & 
\multicolumn{2}{c|}{Apollo 14}	& \multicolumn{2}{c|}{Apollo 15} & 
\multicolumn{2}{c|}{Lunokhod 2} \\\cline{4-11}
 	& diam., m  & years & 
Number&\%-age\footnotemark[1]&Number&\%-age&Number&\%-age&Number&\%-age  \\\hline\hline
Number of CCRs $\times$ size &&& \multicolumn{2}{c|}{100 $\times$ 3.8 cm}& 
                \multicolumn{2}{c|}{100 $\times$ 3.8 cm}&
                \multicolumn{2}{c|}{300 $\times$ 3.8 cm}&
                \multicolumn{2}{c|}{14 $\times$ 6.5 cm} \\\hline
Present/Initial signal strength & &  	& 
                \multicolumn{2}{c|}{$\sim$0.1/1} & 
                \multicolumn{2}{c|}{$\sim$0.1/1} & 
                \multicolumn{2}{c|}{$\sim$0.3/3} & 
                \multicolumn{2}{c|}{$\sim$0.03/$\sim$1} \\\hline\hline 
McDonald  	& 2.7	& 1970-1985	
                & 468 &13.6\%& 495 &14.3\%& 2356 &68.3\%& 132 & 3.8\%\\\hline
Haleakala	& 1.8\footnotemark[2]	& 1984-1990	  
                & 20  &2.9\%& 23  &3.3\%& 633	&91.2\%& 18 &2.6\%\\\hline
OCA    	& 1.5	& 1984-2005	
                & 873 &9.5\%& 774 &8.4\%& 7245	&78.9\%& 285 &3.1\%\\\hline
MLRS\footnotemark[3]  	& 0.76	& 1985-1988	  
                & 10  &3.6\%& 26  &9.5\%& 236	&85.8\%& 3 &1.1\%\\\hline
MLRS\footnotemark[4]	& 0.76	& 1988-2009	
                & 226 &7.9\%& 236 &8.2\%& 2394	&83.5\%& 12 &0.4\%\\\hline
APO	& 3.5 	& 2006-2009	  
                & 121 &18.8\%& 122 &19.0\%& 366	&57.0\%& 33 &5.1\%\\\hline
Total (sum of above): 	&	& 1970-2009	
                & 1718 &10\%& 1676 &9.8\%& 13230 &77.3\%& 483 &2.8\%\\\hline
\end{tabular}
\end{center}
\footnotetext[1]{Percentage of the total number of ranges.}
\footnotetext[2]{Equivalent diameter for multiple lenses in the receiver.} 
\footnotetext[3]{Earlier McDonald Laser Ranging Station (MLRS) site.} 
\footnotetext[4]{Later MLRS site.} 
\caption{Maximum/median/minimum LLR photon rates in photons/minute from each station for each retro-reflector array for period 1984-2009. Rate data is not available for the McDonald 2.7 m site or the first MLRS site. APO is from \cite{Murphy-etal-2009}.  
\label{tab:5-LLR-photon-rates}}
\begin{center}
\begin{tabular}{|c|c|c|c|c|c|c|}\hline 
Station	& Receiver & Time span,  & Apollo 11 & Apollo 14& Apollo 15 & Lunokhod 2 \\
& Diam. m  &   years & max/med/min & max/med/min   &  	max/med/min    & max/med/min  \\\hline\hline
Number of CCRs $\times$ size & &  			& 100 x 3.8 cm 	& 100 x 3.8  cm & 300 x 3.8 cm 	& 14 x 6.5 cm \\\hline
Present/Initial signal strength & &  	& $\sim$0.1/1	& $\sim$0.1/1	& $\sim$0.3/3	& $\sim$0.03/$\sim$1 \\\hline\hline
Haleakala & 1.8	& 1984-1990 &  --	   &-- &4.7/1.3/0.2 &--\\\hline
OCA       & 1.5	& 1984-2005 &  19.0/2.7/0.4 & 26.0/2.7/0.4 & 58./3.5/0.2 &	7.1/1.5/0.4 \\\hline
MLRS & 0.76 & 1988-2009 &  3.3/0.5/0.1& 4.2/0.7/0.1 & 11./0.8/0.1 & 4.3/0.6/0.1 \\\hline
APO & 3.5 	& 2006-2009   & 1070/48/1.4 & 1820/52/1.9 & 3770/45/0.9 &	135./14/1.4 \\\hline 
\end{tabular}
\end{center}
\vskip -10pt
\end{table}
 
Table~\ref{tab:3-LLR-ranges-station} also lists the number of normal points per target as a percentage of the total for a given station. The relative number of ranges from different arrays clearly depends on the size and relative reflected signal strength of the array. The number of ranges to the Apollo 11 and 14 arrays is nearly equal, as expected for their nearly identical designs. If reflected signal strength was not a factor, we would expect similar numbers of ranges from the Apollo 15 and smaller arrays. But we see an order-of-magnitude more Apollo 15 ranges for OCA and MLRS and even a factor of 3 for APO, which collects the most photons of any station. The stations know that if they cannot get ranges on the Apollo 15 array, the others are undetectable, so Apollo 15 is the first array attempted in the night and it is the one attempted when conditions are difficult. 

The tabulated initial return signal strengths for the three Apollo retro-reflector arrays come from the number of corner cubes normalized to the 100 corner cubes of 3.8 cm diameter in the Apollo 11 and 14 arrays. McDonald observatory estimated the initial signal strength of the Lunokhod array, when not exposed to sunlight, as comparable to the two small Apollo arrays. Murphy et al. \cite{Murphy-etal-2009,Murphy-etal-2010} estimate the current signal strength as $\sim$1/3 of the Apollo 11 and 14 strengths. The annual number of Lunokhod observations from OCA and the MLRS has decreased; the MLRS made its last observation in 1996. The Lunokhod array is now a difficult target that can be ranged only by large telescopes on nights with good quality seeing; its fading appears to be real. The Lunokhod corner cube array may be affected by levitated dust and/or ultraviolet degradation of the reflective coating. The Apollo corner cubes do not use reflective coatings, relying on total internal reflection, but they can by affected by levitated dust. Ratios of signal strengths for different arrays are easier to estimate than changes in signal strengths over time because of changes in telescopes, detectors, and laser wavelengths. The estimates of the current signal strengths come from \cite{Murphy-etal-2009} comparing the strongest APO signals with computed values. According to those estimates, the Apollo retro-reflectors are a factor of $\sim$10 weaker than expected for the fresh retro-reflectors and there is significant fading of all reflectors. Levitated dust is suspected as a cause. 

\begin{figure}[!h]
    \begin{center} 
\begin{minipage}[t]{.46\linewidth}
 \epsfig{file=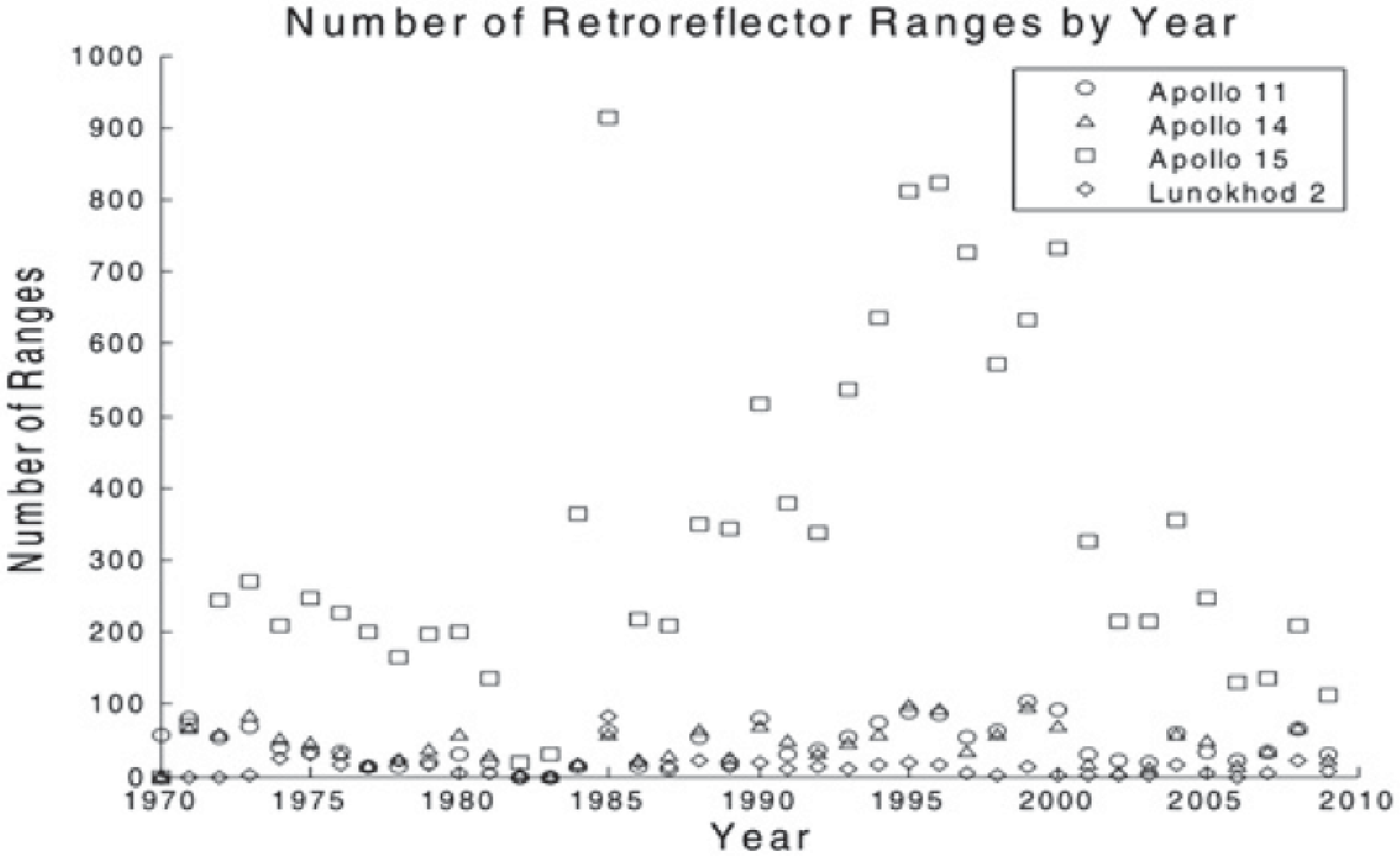,width=80mm} 
  \caption{Annual number of lunar ranges for each of four arrays.}
\label{fig:6-no-photons-by-station}
    \end{minipage}
\hskip 6pt
\begin{minipage}[t]{.45\linewidth} 
\epsfig{file=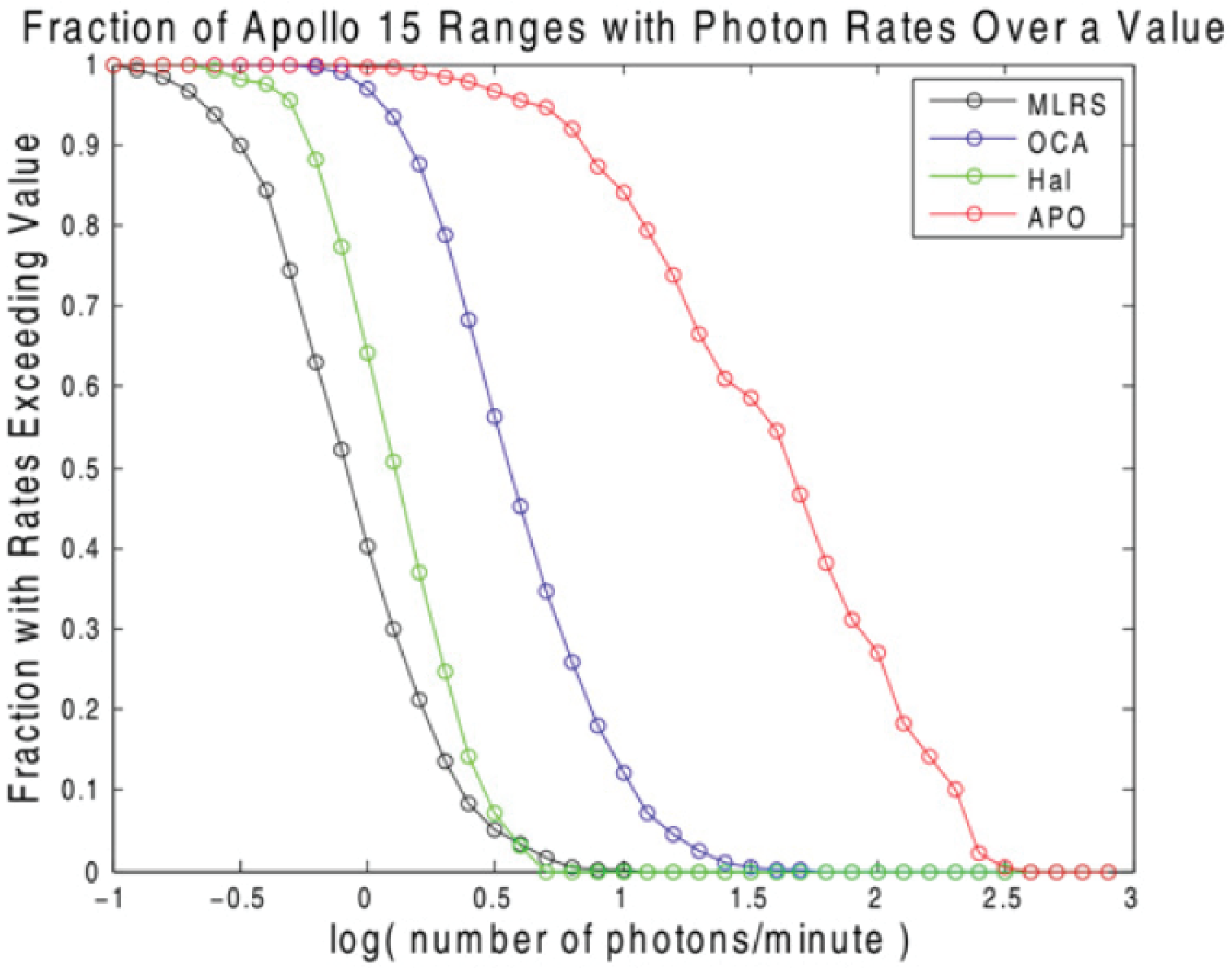,width=63mm}
  \caption{Fraction of Apollo 15 ranges with photon rates exceeding the value of the abscissa.}
\label{fig:7-fraction-A15-ranges}
    \end{minipage}
    \end{center}
  \vspace{-15pt}
\end{figure}

The number of ranges for each retro-reflector array for each year is plotted in Figure~\ref{fig:6-no-photons-by-station}. The dominance of the larger Apollo 15 array over the smaller arrays is clear. Changes in telescopes, detectors, and laser wavelength make it more difficult to see effects of changes of signal strength with time. However, the ratio of the number of Lunokhod 2 ranges to Apollo ranges does decrease with time, lending support to the idea that at least that retro-reflector is fading with time, making it harder to acquire a return signal. 

The detected photon rate can vary greatly depending primarily on atmospheric scintillation. The maximum and minimum photon rates for a station-reflector pair indicate how variable the conditions can be. 

Table~\ref{tab:5-LLR-photon-rates} gives maximum, median and minimum photon rates by station site and array based on the range data set. This information is not available from the active data format for the earliest period. For each station, the minimum rate is fairly consistent for different arrays indicating a threshold for finding the signal. The maximum rate is a better indicator of the strength of the return from each retro-reflector array. OCA and the MLRS maxima show approximately the expected ratios for the Apollo arrays. The OCA maximum for Lunokhod 2 is about 1/3 that of the small Apollo reflectors, as Murphy has estimated using APO. An anomaly is the MLRS maximum rate for the Lunokhod reflector is comparable to the small Apollo arrays; the last of 12 MLRS ranges to that array was in 1996 while MLRS continues to range the Apollo 11 and 14 arrays in the last decade, again implying a currently weak Lunokhod signal. 

The formerly lost Lunokhod 1 array was found by the camera on the Lunar Reconnaissance Orbiter (LRO) spacecraft  in Feb. 2010 and first ranged in April 2010 \cite{Murphy-etal-2011}. Surprisingly, it has strength between the small and large Apollo arrays when in the dark, but weakens when heated by sunlight. It is certain that the Lunokhod 2 array reflected signal strength has faded with time and it is suspected that all of the retro-reflectors are coated with a thin layer of dust that is limiting their performance. 

The distribution of number of observations vs photon rate has also been computed for the four stations. The cumulative distributions for Apollo 15, the fraction of observations with photon rates higher than some value, are shown in Figure~\ref{fig:7-fraction-A15-ranges}. The median values in Table~\ref{tab:5-LLR-photon-rates} come from the 0.5 cumulative fraction. The middle part of each curve is approximately linear on a semilog plot. The upper left side of each curve rolls over as incomplete sampling sets in, and there is a small number of exceptional nights at the lower right end of the curves. We can use the distribution functions to predict the number of ranges as a function of the reflector signal strength. A weaker reflection from the retro-reflector should shift the photon rate lower, and it also alters the fraction since the total changes. On the linear portion of the MLRS and OCA curves for Apollo 15, a factor of 3 weaker signal corresponds to loss of half the data for MLRS and OCA, while for APO, the loss would be closer to 1/3. For a 10 times weaker signal, we cannot stay on the linear portion of the curves for MLRS and OCA and most of the data would be lost compared to Apollo 15, while for APO we would lose $\sim$60\% of the data on the linear portion. A more complete computation would need to account for the nonlinear portion of the curves at the left side where they roll over due to incompleteness. These simple estimates may underestimate the loss of data since Table~\ref{tab:5-LLR-photon-rates} implies even greater sensitivity to signal strength. 

The relative numbers of observations and the photon rates both show the importance of the return signal strength of the retro-reflector arrays. Clearly, for all stations, signal strength affects ranging success. The small Apollo arrays give less data than the larger Apollo 15, and the Lunokhod 2 is a very difficult target.  This leads us to advocate a large single corner cube with reflected signal strength comparable to the original Apollo arrays of small corner cubes.

\section{\label{sec:3} Instrument description}

Our objective was to develop a retro-reflector that gives strong reflected signal strength with little spread of the pulse length. A reflected signal strength at least as strong as the small Apollo arrays was set as a minimum design goal. Another important goal was to maximize the fraction of the lunar day that the device was usable, driving the thermal performance. To reduce the pulse length spread we opted for a single corner cube approach, eliminating active alignment designs as impractical. This choice together with the strong signal requirement drove us to a special diffraction pattern design that covers the aberrated position of the Earth station.

\subsection{\label{sec:2.3.1} Aberration}

The aberration problem with a large reflector can be understood by considering its diffraction pattern. For incoming light normal to the front face, the diffraction pattern of an ideal circular corner cube consists of a central spot surrounded by rings if the corner cube has reflective coatings, or six spots, if total internal reflection is used \cite{Chang-etal-1972,Arnold-2002,Otsubo-etal-2010}.  For illumination off normal incidence, these patterns are elongated and distorted.  The Apollo corner cubes have a diffraction pattern with a central spot within $\sim$3 arcsec radius and a ring of six spots farther from the center for normal incidence at the 532 nm wavelength currently in use \cite{Arnold-2002,Arnold-2004}.

\begin{figure}[!h]
    \begin{center} 
  \vspace{-8pt}
\begin{minipage}[t]{.46\linewidth}
 \epsfig{file=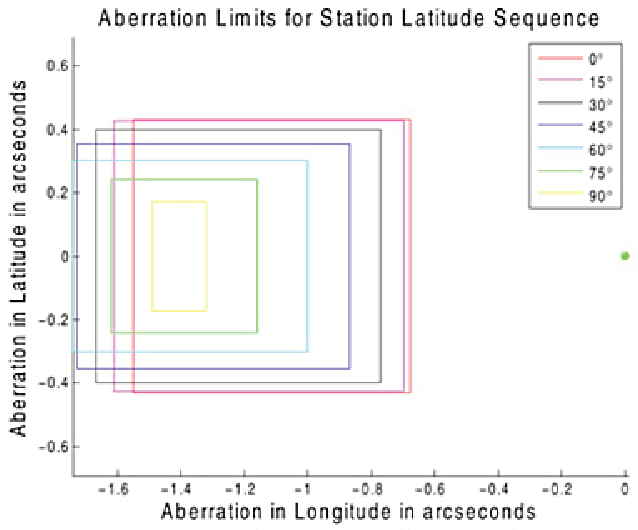,width=62mm} 
  \caption{The green spot at the origin marks the direction of the laser beam arriving at the Moon. The reflected diffraction pattern must include the boxes marking the limits of aberration for stations at different latitudes.}
\label{fig:8-LLR-aberration-limits}
    \end{minipage}
\hskip 10pt
\begin{minipage}[t]{.45\linewidth} 
\epsfig{file=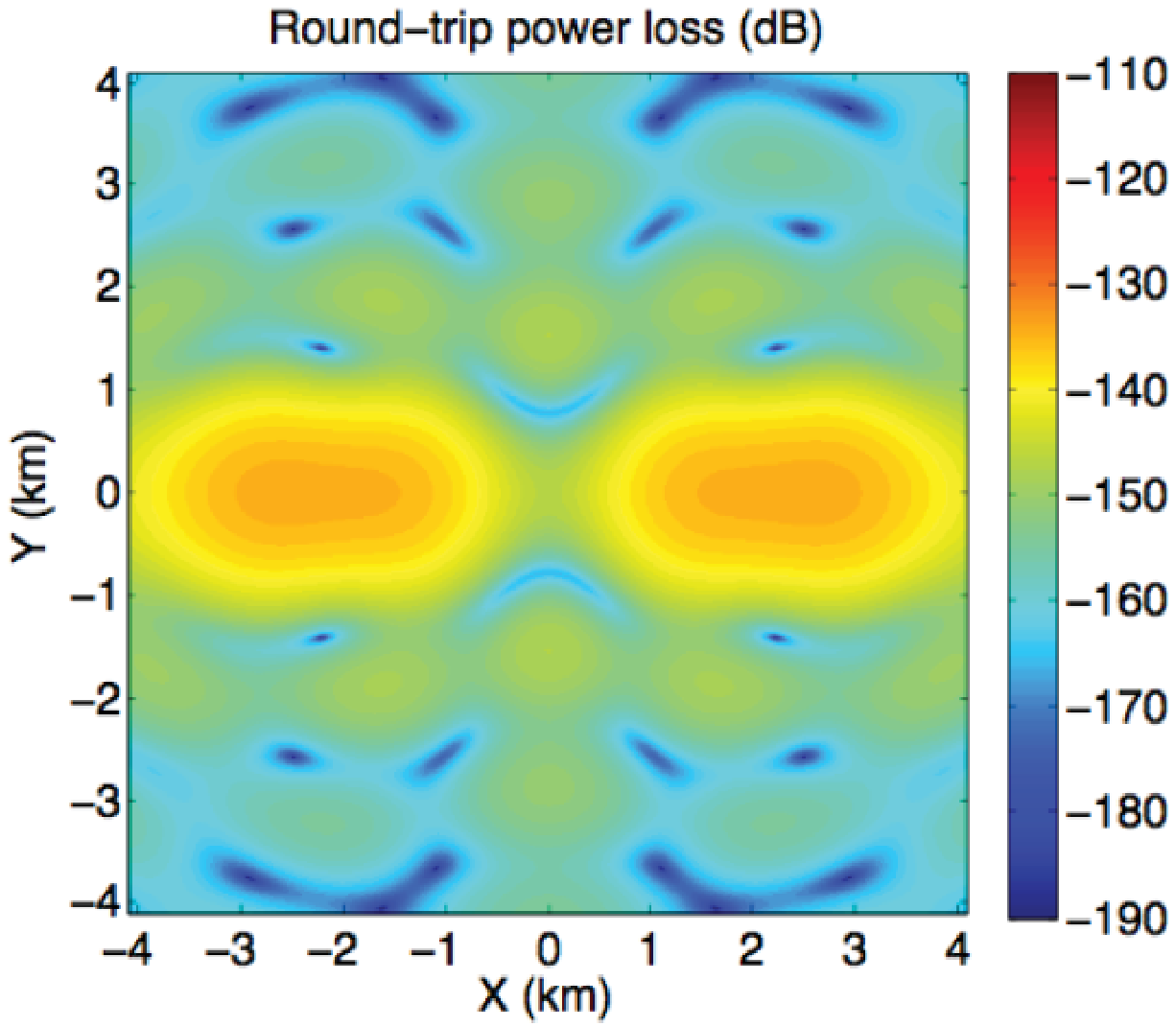,width=58mm}
  \caption{Diffraction pattern split in one dimension. 1 arcsec matches 1.9 km at the mean lunar distance and one lobe of the pattern coincides with the boxes of Fig.~\ref{fig:8-LLR-aberration-limits}.}
\label{fig:15-CCR-diffraction-1D}
    \end{minipage}
    \end{center}
  \vspace{-10pt}
\end{figure}

An ideal CCR reflects light back to the point it originated from, but for LLR the tracking station moves between the transmission and reception times, due to the $\sim$1.0 km/sec motion of the Moon in its orbit and the rotation of the Earth (465 m/sec at the equator).  The pattern of aberration is shown in Figure~\ref{fig:8-LLR-aberration-limits}. The green spot at the origin is the apparent location of the incoming laser beam.  As seen from the Moon, the apparent position of the station on Earth moves by an angle in the range 0.7--1.7 arcsec (or 1.3--3.2 km) during the 2.5 sec round-trip travel time of the laser pulse between transmission and reception.  The existing arrays of small 3.8 cm Apollo corner cubes accommodate this velocity aberration with a large beam pattern with a central spot size that exceeds the largest aberration without any special design effort.  A larger CCR with a smaller beam spread needs a carefully designed diffraction pattern to encompass the expected range of velocity aberration. 

In our instrument this problem is solved by modifying the shape and direction of the return beam which is achieved by altering the dihedral angles by small amounts to compensate for the aberration effect.

\subsection{\label{sec:3.geom}	Orientation of the instrument}

A lunar CCR instrument should be aligned so that the normal to the front face of the array points toward the mean Earth direction, the center of the pattern in Figure~\ref{fig:15-CCR-diffraction-1D}.  Failure to properly orient an array will broaden the return pulse.  The orientation of the Apollo retro-reflectors was adjusted by the astronauts.  Apollo 11 had a sighting device toward the Earth.  Apollos 14 and 15 each had a bubble level and a device casting a shadow from the Sun.  The orientation was adjusted until the bubble indicated level and the shadow matched a mark.  This worked well, roughly $\pm 1^\circ$ accuracy for pointing.  The Lunokhod arrays were rigidly attached to the rover and were pointed by moving the vehicle.  Their orientation accuracy was presumably worse. Therefore, a retro-reflector instrument must have some means for orienting the normal to the front face toward the mean Earth direction. In the case of a single CCR, which is not affected by the lunar libations, a proper orientation would maximize the returned signal. 

In our instrument the diffraction pattern is split into two spots (Figure~\ref{fig:15-CCR-diffraction-1D}). This split is done either by altering one of the dihedral angles by 0.8 arcseconds from 90$^\circ$ or with a corrector plate combined with the solar filter. The axis of one-dimensional split (i.e., the line joining the two spots) must be aligned with the mean aberration direction parallel to the lunar equator; hence the instrument must be oriented correctly when deployed.

\subsection{\label{sec:2.3.2} Thermal Effects}

The lunar environment presents major challenges to the design and performance of any optical system.  Temperatures may range from a high of 390 K during the lunar day to a low of 90 K during the lunar night.  Sunlight will come from a variety of directions as the Sun moves slowly across the sky during the day. There is no atmosphere to moderate thermal differences. The heating from sunlight puts stringent limits on arrays for operation during the lunar day. A thermal gradient across a solid corner cube prism will cause a gradient in the index of refraction which leads to spreading of the angle of the return beam, giving a much larger beam diameter on return to the Earth, thus weakening the signal strength at the receiving station.

Daylight operation is desired and thermal effects force several design compromises and conditions.  Thermal gradients in solid prisms must be minimized so the prism size is limited, a reflecting coating cannot be used on the back faces, and protection from sunlight is desirable. An alternative to solid prisms, open corner cubes have thermal benefits. 

The current reflector arrays on the Moon use solid glass CCRs. These provide good dimensional stability, can use total internal reflection, and are easier to manufacture than a hollow CCR.  However, a large solid CCR will have a much larger mass, so a hollow CCR is preferred to ease launch and robotic deployment requirements.  A hollow CCR must use coated surfaces for signal reflection. Temperature gradients within the material of a hollow CCR will lead to mechanical distortions of its surfaces, which will also have a deleterious effect. 
Thus, control of thermal effects within the instrument is important, and front surface mirrors are probably needed. 

To counteract thermal effects the Apollo arrays used small prisms (3.8 cm across), recessed the prisms 1.9 cm in an aluminum plate, and relied on total internal reflection rather than using reflecting coatings on the prism's back faces. The aluminum plate is a good thermal conductor and will have a nearly uniform temperature. Solid cube corners that are partly in sunlight and partly in shadow have thermal gradients leading to weaker signals. Larger prisms are more sensitive to thermal gradients. Reflecting coatings on the back faces (silver or aluminum) heat up in sunlight causing temperature gradients and spoiling the beam pattern. The French CCRs on the Lunokhods are silvered; their return signal drops drastically during lunar daylight \cite{Fournet-1972}.

To mitigate the impact of thermal effects, our instrument relies on an enclosure to shield the CCR from direct sunlight.  The enclosure uses high thermal conductivity aluminum to minimize temperature gradients in the shell. In order to minimize thermal gradients in the CCR due to absorbed sunlight, a narrow-band optical filter is used at the input of the CCR to pass the chosen laser-ranging wavelength (532 nm), while blocking most of the sunlight.  The solar filter will be made from precise optical coatings applied to a fused silica substrate, which can also serve as a phase corrector plate.  There is also a multi-layer insulation (MLI) inside the shell to minimize optical distortions due to thermal gradients across the CCR. Combination of these factors leads to thermal gradients resulting in only small fairly symmetric distortions (see Fig.~\ref{fig:18-CCR-simulated-thermal} below). An optical analysis of the thermally induced distortions shows an acceptable change to the beam patterns (discussed in Section~\ref{sec:2.5} below).

\subsection{\label{sec:2.3.3} Lunar Environment}

After nearly four decades on the Moon the Apollo retro-reflectors continue to give useful return signals, although apparently weaker due to long-term degradation. The performance of the arrays appears to have degraded since the beginning of measurements \cite{Murphy-etal-2010}. Only a handful of terrestrial laser ranging facilities are capable of detecting the weak signals and making the LLR measurements. The charge on the lunar surface changes between day and night.  It has been suggested that this causes dust grains to levitate.  This would be a concern for any optical device on the Moon including retro-reflectors. 

The Lunokhod 2 retro-reflector initially gave as strong a return signal as the Apollo 11 and 14 arrays, even after traveling more than 20 km across the lunar surface.  Modern stations report a much weaker return signal than for the Apollo retro-reflectors.  This might be due to dust (the Lunokhod corner cubes are more exposed since they are not recessed) or a degradation of the silver coating on the corner cube back reflecting faces due to the harsh environment or darkening of the glass under UV radiation.

We emphasize that the CCR must remain dust free and be mounted on a stable platform. If a robotic deployment is realized, venting of all spacecraft tanks is required shortly after landing. Our instrument includes a full-aperture dust cover placed over the CCR container which will be removed as the last step in the deployment sequence. During the science operations,  the instrument will rely on electrostatic dust mitigation techniques, which are currently being investigated for this purpose.  

\subsection{\label{sec:3.1} Hollow corner cube or solid prism approach?}

An obvious design consideration is the choice for the structure of the corner cube -- should it be an open or solid one? It is clear that a larger solid CCR would have more mass than a comparably sized hollow CCR.  Thus, a 100 mm solid CCR \cite{Currie-etal-2009,Currie-etal-2011} will have mass comparable with our 170 mm hollow CCR.  The 100 mm solid CCR would give signal return comparable to the current (degraded) performance of the Apollo 15 CCR array.  

\begin{wraptable}{R}{0.45\textwidth}
\vskip-10pt
\caption{Return signal strength for single CCRs normalized to ideal Apollo 11/14 arrays. Apollo 15 would have strength 3. A diffraction pattern spread in one or two dimensions is indicated by 1D or 2D. Uncoated solid CCRs use total internal reflection. Coated ones have a reflecting coating on the back faces. The hollow CCRs have a reflecting coating. The new 17 cm and the UMd/INFN-LNF 10 cm CCRs \cite{Currie-etal-2009,Currie-etal-2011} are in bold.
\label{tab:6-return-strength}}
\begin{center}
\vskip -5pt
{\small 
\begin{tabular}{|c|c|c|c|c|}\hline
Corner cube type	& 10 cm	& 15 cm	& 17 cm	& 20 cm\\\hline\hline
Solid, uncoated	& 0.2	& 0.8	& --	& 1.6, 2D\\
Solid, coated 	& 0.2	& 0.7, 2D& --	& 1.5, 2D\\
Hollow 	& 0.2	& 0.7, 2D	& 3, 1D	& 1.5, 2D\\\hline
\end{tabular}
}
\end{center}
\vskip -10pt
\end{wraptable}

Table~\ref{tab:6-return-strength} gives the return signal strength expected from several ideal corner cube types and sizes normalized to the ideal array of 100 of the 3.8 cm corner cubes of Apollo 11 and 14. Apollo 15 would have signal strength 3. 
The 17 cm hollow CCR is the design proposed here and 1D refers to a one-dimensional split of the diffraction pattern. The other 9 cases are from \cite{Otsubo-etal-2010}. The 10 cm solid case corresponds to the UMd/INFN-LNF corner cube \cite{Currie-etal-2009,Currie-etal-2011} is an existing concept to build single CCR utilizing a solid prism approach. Currie et al.~\cite{Currie-etal-2011} estimate that the 10 cm UMd/INFN-LNF design would have signal strength 0.25 compared to the original small Apollo arrays, a little larger than Otsubo et al. estimated. At 17 cm our one-dimensional-split design outperforms the two-dimensional design by a factor of $\sim$3. The 17 cm design outperforms the 10 cm solid by a factor of at least 12, making it an easy target of modern LLR stations.

It is not clear whether the Apollo arrays have faded significantly over 4 decades. To allow for this uncertainty in our design, we considered the following two possible extremes: 
\begin{inparaenum}[i)]
  \item If the Apollo arrays have not faded, then Table~\ref{tab:3-LLR-ranges-station} indicates that the 17 cm design would give a signal strength for existing stations that would gather data at a rate comparable to Apollo 15. The 10 cm solid design would have a 15 times weaker signal and would be comparable to, or weaker than, the Lunokhod 2 array. 
  \item If the Apollo arrays have faded by a factor of 10, then the 17 cm design will give a booming signal when new, which might be detected by smaller laser ranging stations, and the 10 cm design would be somewhat weaker than the present Apollo 15, but still very good. If levitated dust or some other factor has degraded the Apollo arrays, then new reflectors may also fade with time. The desired longevity of use argues for a strong return signal from the new instrument.
\end{inparaenum}

Our instrument relies on a single large hollow CCR to yield a 1-mm class Earth-Moon range precision in a single pulse, while still keeping a low mass, allowing for a robotic deployment.    

\subsection{\label{sec:3.2} Choice of materials and design optimization}

Ideally, one would like to make the CCR out of a material that would have a zero coefficient of thermal expansion (CTE) in the temperature range 50--350~K; however such a material does not exist.
 
Fused silica was used for the Apollo CCRs and it is an excellent material.  The four faces of the prism must be of high optical quality and the angles between the reflecting faces should be within a fraction of an arcsec of 90$^\circ$ to prevent distortion of the diffraction pattern.  Fused silica has a low CTE, $5\times10^{-7} ~/^\circ$.  The refractive index is 1.455 and its sensitivity to temperature is $1\times10^{-5}~/^\circ$.  Thermal gradients across the CCR are a limitation for daytime use and the sensitivity of the refractive index to temperature is more important than the thermal expansion coefficient. Faller \cite{Faller-1972} estimates that fused silica prisms larger than about 4.7 cm diameter will suffer significant thermal degradation during the day and that smaller CCRs are required for near diffraction limited performance during lunar day.  Laboratory tests confirmed the essential correctness of the Apollo thermal design \cite{Chang-etal-1972}. 

Given real-world materials with non-zero CTE and the need to operate over a large temperature range, our approach was a) to build a CCR in as uniform a fashion as possible, and b) to keep the CCR's temperature as uniform as possible. This means that, as the temperature changes, the CCR may expand and contract uniformly, preserving the angles and flatness of the optical surfaces. The CCR can tolerate the dimensional changes of $\sim$100 microns, but one must constrain changes in the CCR angles to be less than 0.1 arcsec. This leads to our design that can be summarized as having: a) a thermally-conductive container, b) a multi-layer insulation, and c) a solar filter. This was all intended (and shown by modeling, see Section~\ref{sec:2.44}) to result in a uniform temperature for the CCR within the range of $0.1$~K (as it was varied in the range 50--350~K) with minimal change in the returned light pattern. 

Based on the above, the most desirable characteristics of the CCR material for our instrument are: i) a bonding method which does not introduce a material layer with a CTE different from the face material, ii) low and uniform CTE in the CCR material, ii) a CTE which matches the CTE of the reflective coating, iv) high stiffness, v) high thermal conductivity, vi) low thermal mass, vii) low density, viii) high-quality, optically-flat surfaces, and ix) meets all the requirements of a space-qualified material. Although silica may not be the final choice, selecting Zerodur or ULE on the basis of lower CTE is not necessarily the best way to go. A lower CTE may help somewhat, but is not the only consideration. At present our tentative choice is silica.

\subsection{\label{sec:3.sum} Summary of the design}

\begin{wrapfigure}{R}{0.36\textwidth}
 \vspace{-15pt}
  \begin{center}
    \includegraphics[width=0.36\textwidth]{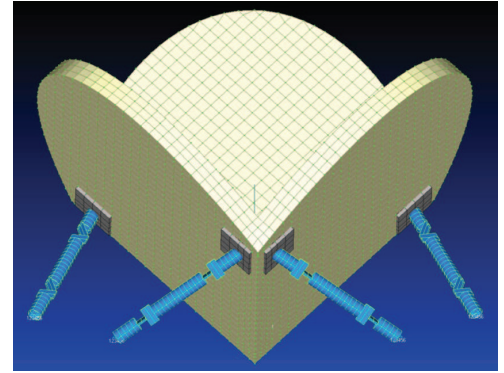}
  \end{center}
  \vspace{-10pt}
  \caption{
A large hollow CCR. The face shape is chosen so as to result in a circular aperture for the CCR (for illustration, see left part of Fig.~\ref{fig:18-CCR-simulated-thermal}). This choice leads to a CCR with negligible pulse spread and the best aperture-to-mass ratio over other shapes.}
\label{fig:14-LLR-large-CCR}
  \vspace{-5pt}
\end{wrapfigure}

The trade-off analysis presented above resulted in the design which is based on a combination of the following new features: i) a large silica-based hollow CCR with a deliberately introduced small dihedral angle offset, ii) a solar filter with a narrow optical band-pass, iii) a cylindrical casing, and iv) thermal insulation. The dihedral angles need to be correct to within 0.8 arcsec to achieve the desired beam pattern.  This is near the current capability for assembling CCRs.  To allow for some relaxation in assembly tolerances, our design allows for the inclusion of a phase corrector plate at the CCR opening. After assembly of the CCR, the reflected beam pattern will be measured using an optical interferometer, and the phase corrector plate will be designed and manufactured to result in the desired beam pattern.

The present version of our CCR is shown in Fig.~\ref{fig:14-LLR-large-CCR}.  Its optical aperture diameter is 170 mm, about five times the 38 mm of the small CCRs in the Apollo arrays.  The three facets comprising the cube are shaped to create a circular optical aperture for normal incidence light.  There are four commonly-used shapes for the faces of a CCR: triangular, square, clipped square (or pentagonal), and a face shape which results in a cylindrical shape for the entire CCR.  The choice of a cylindrical CCR aperture maximizes the optical return while minimizing the mass.  

This design is new, it successfully addresses all the identified problems, allows building an instrument capable of high Earth-Moon ranging precision of $\leq$~1~mm, while also having low mass (i.e., $\leq$~3.5~kg total mass including the cube, filter, support, casing and insulation), allowing for robotic deployment and drawing no power.

\section{Modeling and testing of the new instrument}
\label{sec:model-test}

\begin{wrapfigure}{R}{0.49\textwidth}
 \vspace{-10pt}
  \begin{center}
    \includegraphics[width=0.48\textwidth]{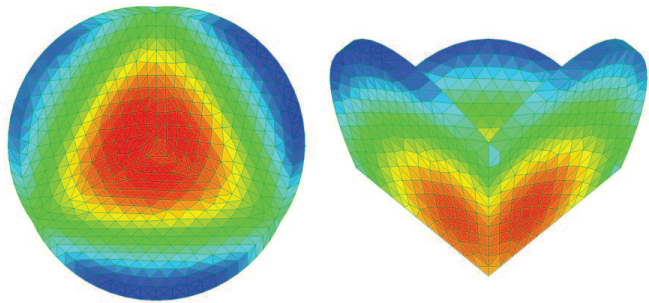}
  \end{center}
  \vspace{-10pt}
  \caption{Simulated thermal performance of the CCR instrument under lunar day conditions. Left: top view of the CCR; right: angled view of the CCR. Entire instrument was modeled, including the corner-cube retro-reflector, narrow band-pass filter, the canister, bipod mounts, as discussed in Sec.~\ref{sec:2.44}. Temperature range: blue $-39.44^\circ$~C, red $-32.32^\circ$~C.}
\label{fig:18-CCR-simulated-thermal}
  \vspace{-0pt}
\end{wrapfigure}

\begin{figure}[!h]
    \begin{center} 
\begin{minipage}[t]{1.0\linewidth}
 \epsfig{figure=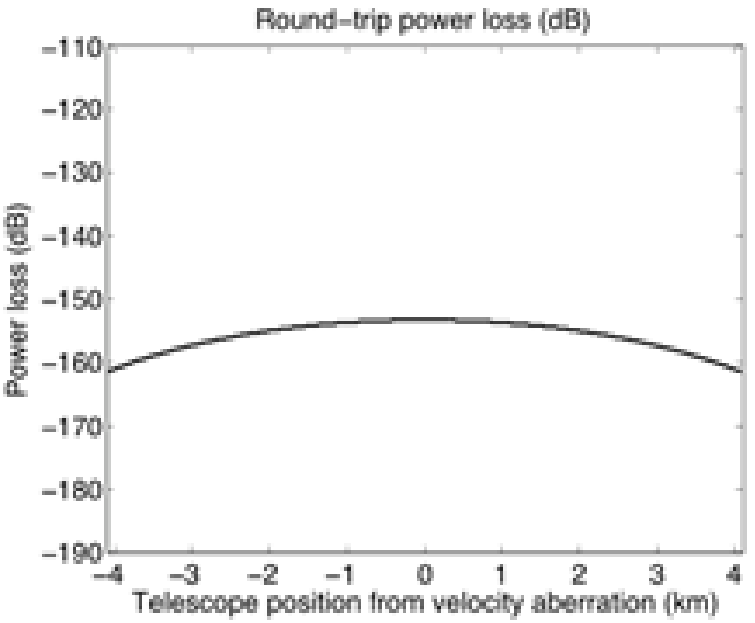,width=52mm} 
\hskip 0pt
\epsfig{file=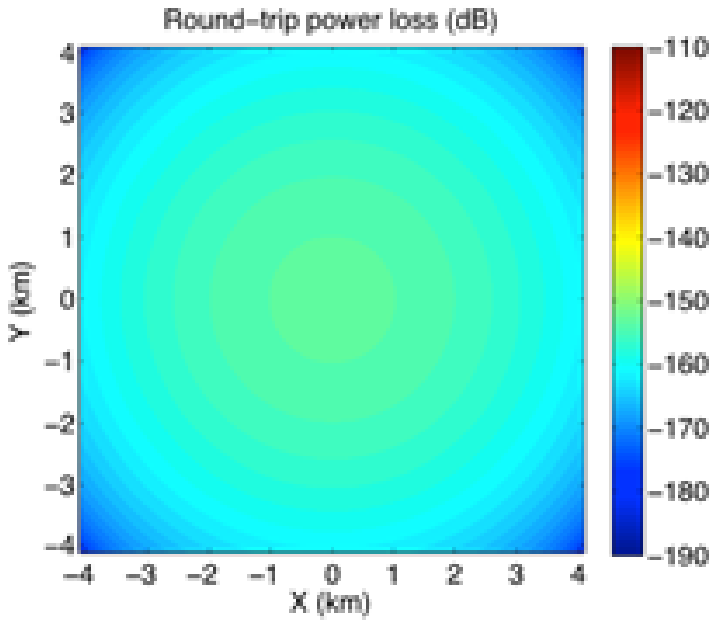,width=51mm}
    \end{minipage}
\vskip -5pt
\caption{Round trip attenuation for a single, 38mm diameter, Apollo-sized CCR.} 
\label{fig:9-round-trip-pow-loss}
\vskip -15pt
    \end{center}
    \begin{center} 
\begin{minipage}[t]{1.0\linewidth}
 \epsfig{figure=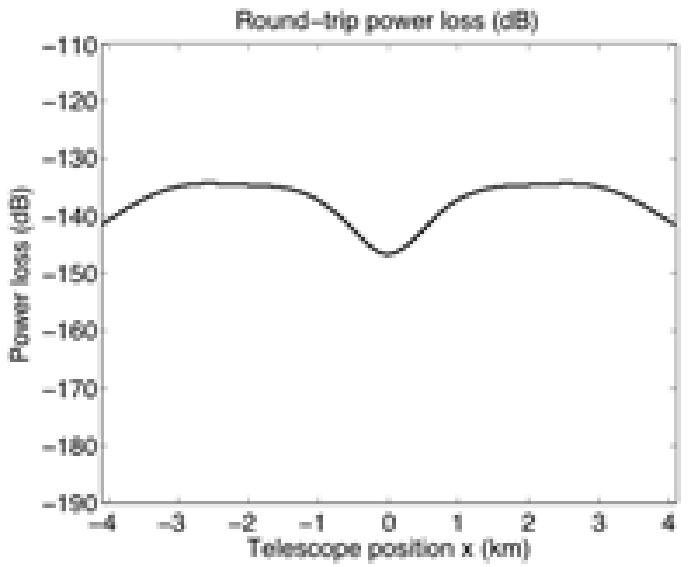,width=53mm} 
\hskip 0pt
\epsfig{file=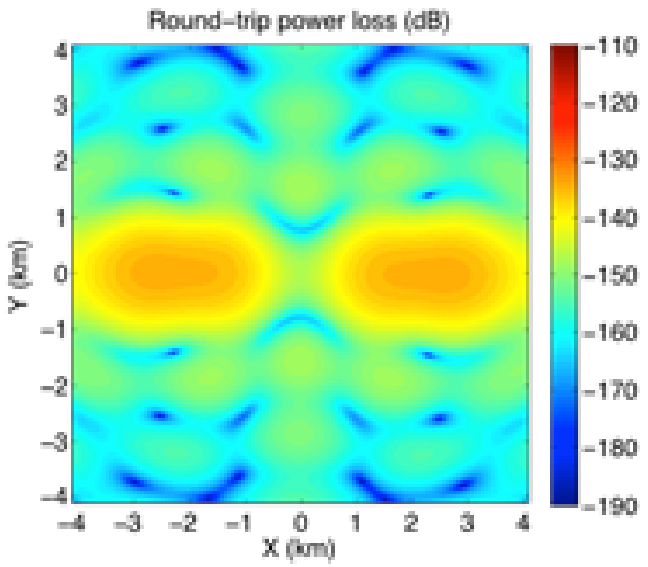,width=51mm}
    \end{minipage}
\vskip -5pt
\caption{Receiving telescope patterns with thermal distortion caused by heating from the Sun at a 35.3 degree angle of incidence (worst case with one CCR face illuminated).} 
\label{fig:11-round-trip-pow-loss-35deg}
\vskip -15pt
    \end{center}
\vskip -15pt
    \begin{center} 
\begin{minipage}[t]{1.0\linewidth}
 \epsfig{figure=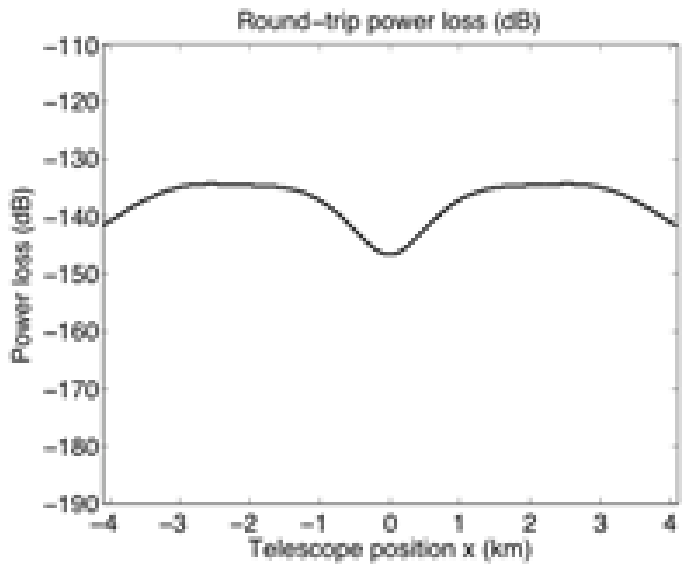,width=53mm} 
\hskip 0pt
\epsfig{file=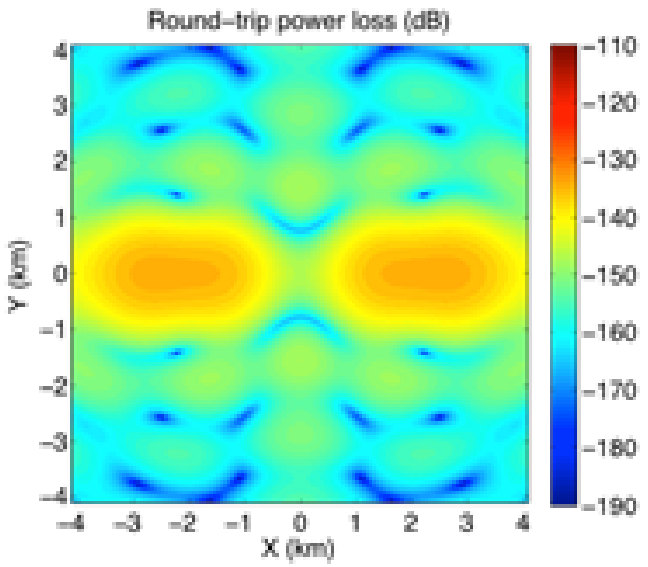,width=51mm}
    \end{minipage}
\vskip -5pt
\caption{Receiving telescope patterns with no thermal distortions.} 
\label{fig:12-round-trip-pow-loss-no-thermal}
    \end{center}
    \begin{center} 
\begin{minipage}[t]{1.0\linewidth}
 \epsfig{figure=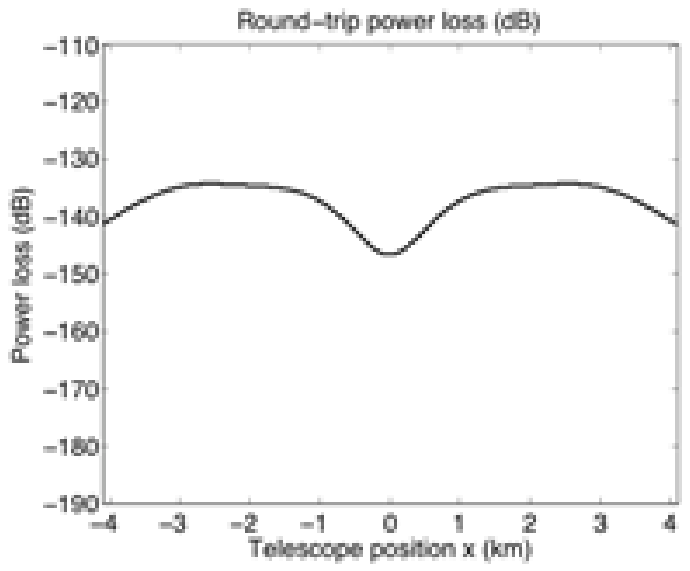,width=53mm} 
\hskip 0pt
\epsfig{file=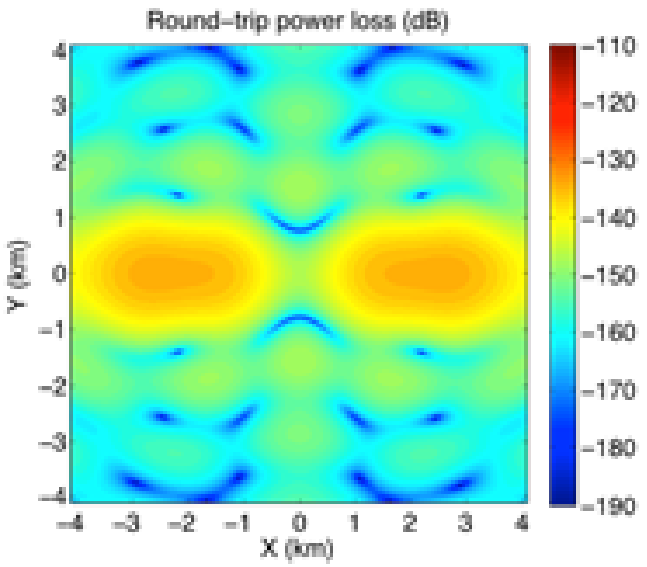,width=51mm}
    \end{minipage}
\vskip -5pt
\caption{Receiving telescope patterns with thermal distortion caused by heating from the Sun directly overhead.} 
\label{fig:13-round-trip-pow-loss-Sun-overhead}
    \end{center}
\vskip -15pt
\end{figure}

After the design requirements were identified, a 170 mm diameter, hollow CCR system was 
tentatively selected and analyzed using a combination of thermal, mechanical, and optical diffraction models. Detailed design and simulations were done to show that the CCR will meet the requirements. A structural-thermal-optical-performance analysis was completed to establish performance for different solar heating conditions. The mechanical design is shown in Figure~\ref{fig:14-LLR-large-CCR}. A thermal distortion analysis result is shown in Figure~\ref{fig:18-CCR-simulated-thermal}. For reference, the round trip attenuation for a single Apollo-sized CCR is shown in 
Figure~\ref{fig:9-round-trip-pow-loss}. Figures~\ref{fig:11-round-trip-pow-loss-35deg}-\ref{fig:13-round-trip-pow-loss-Sun-overhead} show the expected optical pattern optimized to concentrate signal at the transmission and reception locations of the ground station. 

\subsection{\label{sec:2.44} Modeling thermal, mechanical, and optical performance}
 
The modeling routines involved three main steps:
\begin{inparaenum}[i)]
  \item Calculate the temperature distribution throughout the CCR, its enclosure, and solar filter, taking into account the effects of thermal insulation.
  \item Use the resulted temperature distribution to calculate the surface distortion for each of the three faces of the CCR.
  \item Use the face distortion as input to an optical diffraction model to calculate the distribution of laser light upon return to the Earth, while also accounting for the round-trip attenuation factors.
\end{inparaenum}

For the development of the optical model of the instrument  we relied on a computational diffraction model developed at JPL for the Space Interferometry Mission (SIM) Project \cite{Dutta-Benson-2003}. In addition, the instrument model takes into account propagation losses for the round trip from the Earth to the Moon and back, the loss due to the optical aperture area of the lunar CCR, and the loss due to the optical aperture area of the receiving telescope. Attenuations due to Earth's atmosphere and reflection losses at the surfaces of the CCR were not included.

It was assumed that the entire CCR system was positioned on a flat lunar surface with no varying topography near-by (mountains and craters) to influence thermal view factors.  The analysis accounted for the fact that it takes 14 days between lunar sunrise and lunar sunset which justified the reliance on the steady state conditions. The model was overall radiation dominant, with space being treated as 2.7~K boundary condition. The only conductive paths were between the supporting bipods fixing the CCR to its container and a bolted path between the solar filter and the container.  Remaining thermal interactions were radiation based heat transfer. Due to thermal equilibrium the model assumed that lunar surface temperature under container follows container's bottom temperature. Also, there was no heat flow from sunlit lunar surface to covered lunar surface due to low soil thermal conductivity of $\sim$ 0.0015~W/m\,K. The reflector surfaces, phase corrector, and solar filter were modeled as ideal when in thermal equilibrium. The analysis also assumed that there is no lunar dust in the CCR assembly. 

The initial design used for modeling and feasibility studies consisted of a single, open corner cube, located in a cylindrical case.  The transmitting and receiving telescope on Earth was assumed to have an aperture diameter of 1m.  The transmitted wave front was assumed to be Gaussian with a beam diameter of 0.8~m and a wavelength of 532 nm.  The distance from the telescope to the retro-reflector is 384,400 km. A thermal finite element model was developed in I-DEAS TMG\footnote{For details on thermal analysis package I-DEAS TMG, see {\tt http://www.mayahtt.com/}}  that consisted of 2,181 nodes and 3,145 elements. 

The initial design was quickly modified based on thermal constraints.  The first generation model was simply a CCR set within a container.  Both a worst case hot (WCH) and worst case cold (WCC) conditions were established.  The WCH configuration had the sun-facing two of the faces with the third face completely shaded, which resulted in a 35.3$^\circ$ angle from the zenith.  The WCC configuration was no sunlight.  The thermal gradient issue was non-existent for the WCC case since the thermal environment was uniform due to the lack of sunlight; the only open issue was the cold temperatures the CCR experienced, which would be a materials issue to consider for the bondlines.  However, the WCH configuration's adverse solar illumination caused large temperature gradients within the CCR, the sunlit faces of the CCR being hotter than the shadowed face as expected.  The first modification was to limit the incident solar flux on the CCR by incorporating a solar filter.  This reduced the overall temperature of the CCR and reduced the gradient within the CCR as well, but not to the desired level.  Temperature maps showed that the shadowed face was now warmer than the sun-facing sides.  This was due to the CCR picking up heat from the sunlit side of the container.  To mitigate this effect, MLI was wrapped around the container.  This final configuration has less than 0.1~C change between a direct solar load (Sun at zenith) and the worst case angled (35.3$^\circ$) solar load.  

In order to evaluate the thermal distortion, we also constructed a finite element structural model of the corner cube. The mesh had three elements spanning the 10 mm thickness of the glass.  5,319 of 8-node CHEXA\footnote{CHEXA is a six-sided solid element with eight or twenty grid points.} elements were used along with 108 of 6-node CPENTA\footnote{CPENTA is a five-sided solid element with six or fifteen grid points.} elements.  The majority of the model contained CHEXA elements with edge lengths of $5\times 5 \times 3.3$ mm with orthogonal faces.  The model had three constraints located at the outer edge of each vertex representing an ideal kinematic mount.  The constraints, set in a cylindrical coordinate system, limited the vertical and tangential degrees of freedom.  A lookup table was used in order to capture the coefficient of thermal expansion variation with temperature for the glass.  A static analysis solution was found with NX NASTRAN 6.1\footnote{For details on finite element analysis package NX NASTRAN, see {\tt http://www.plm.automation.siemens.com/en\_us/products/nx/nx7/index.shtml}}. 

Temperatures were mapped from the thermal model to the structural model nodes.  To generate the optical surface displacements, the corner-cube was assigned an initial temperature of 22~C and the final temperature was set to the values of the mapped temperatures from the thermal model.  Displacement mapping was not necessary between the optical and structural model due to a one-to-one relation of nodal locations. 

For the results presented in this paper the effects of the CTE mismatch between the glass and the mounting bond pads along with the adhesive was not accounted for.  The stiffness of a mounting structure such as bipods was also not accounted for; as stated earlier the constraints were modeled as perfect kinematic mounts.  The effect of gravitational sag was also left out of this analysis. The remaining set of modeling assumptions closely matched the realistic conditions on the Moon used to support our numerical simulations performed.   

\subsection{\label{sec:2.5} Results of modeling analysis}

As mentioned earlier, the CCRs used in the Apollo retro-reflector instruments were solid corner cubes with a diameter of 38~mm.  For reference, the round-trip attenuation for a single Apollo corner cube was modeled and is shown in Figure~\ref{fig:9-round-trip-pow-loss}.  The attenuation factors for the entire arrays are obtained by adding 20~dB for the Apollo 11 and 14 arrays and 24.8 dB for the Apollo 15 array. Notice that this CCR with perfect dihedral angles creates a large circular diffraction pattern on Earth, which covers the velocity aberration range.

Figure~\ref{fig:14-LLR-large-CCR} shows the modeled design that consists of a 170 mm diameter ``cylindrical'' hollow corner cube made of fused silica 10 mm thick that is enclosed in a 5 mm thick cylindrical aluminum container with an optical filter designed to minimize temperature gradients across the CCR. 

The front of the CCR is protected from direct sunlight by an optical filter, which transmits light in a bandwidth of 10 nm about the 532 nm (frequency doubled Nd-Yag) laser wavelength that is commonplace for ranging.
For comparison, the round trip attenuation for a single Apollo-sized CCR is shown in 
Figure~\ref{fig:9-round-trip-pow-loss}.
Three cases are presented for normal incidence:
\begin{inparaenum}[i)]
  \item The perfect case of no temperature gradients in the corner cube.
  \item The case of the Sun directly overhead.
  \item The worst case of the Sun at an angle of incidence of 35.3$^\circ$. 
\end{inparaenum}
Results are shown in Figures~\ref{fig:11-round-trip-pow-loss-35deg}, \ref{fig:12-round-trip-pow-loss-no-thermal}, and \ref{fig:13-round-trip-pow-loss-Sun-overhead}.

It can be seen that the optical power pattern on return to Earth is virtually unaffected by the presence of sunlight either directly overhead or at an incident angle of 35.3$^\circ$.  The design of the solar filter and the insulated case creates a very good, uniform temperature thermal environment for the CCR.  The optical power returned to the ranging station by this design was shown to exceed the power returned by 100 Apollo solid CCRs.

\subsection{\label{sec:3.4} Testing the new CCR}

\begin{figure}[!h]
  \vspace{-2pt}
    \begin{center} 
\begin{minipage}[t]{.49\linewidth}
 \epsfig{file=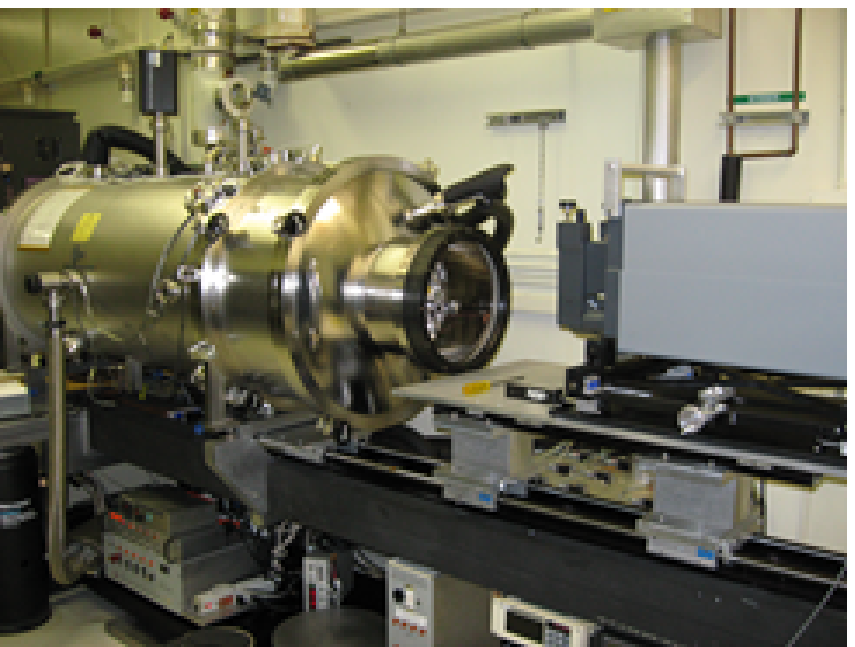,width=55mm} 
  \caption{JPL's thermal-vacuum optical test system with vacuum chamber, large, high quality optical port and optical interferometer.}
\label{fig:19-JPL-thermo-vac-chamb}
    \end{minipage}
\hskip 6pt
\begin{minipage}[t]{.49\linewidth}
\epsfig{file=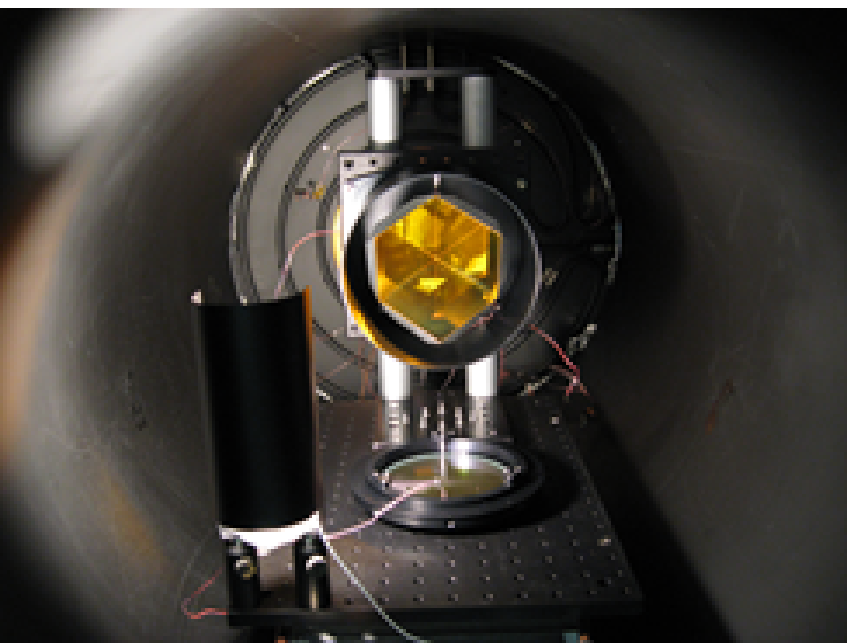,width=55mm}
  \caption{Test of lunar CCR instrument. To show the CCR, the solar filter has been removed and placed below the case. The solar simulator light source is on the left.}
\label{fig:20-JPL-thermo-vac-chamb-CCR}
    \end{minipage}
    \end{center}
  \vspace{-12pt}
\end{figure}

To validate the modeling process a smaller, 125 mm diameter, hollow CCR was borrowed 
from the SIM project at JPL.  The ability of a CCR with reflecting coatings to maintain operation in sunlight through use of a narrow-pass filter was tested with arrangement shown in Figures~\ref{fig:19-JPL-thermo-vac-chamb}, \ref{fig:20-JPL-thermo-vac-chamb-CCR}.  The prototype was placed in a thermal vacuum chamber at JPL to test the entire CCR instrument in a vacuum environment with variable solar illumination. The entire system was modeled using a combination of the same thermal, mechanical, and optical diffraction models used for the large reflector work.

Experimental conditions used in this test can be characterized as follows: 
1) The temperatures of the hollow CCR components were measured (accurate to 1$^\circ$C) using type $T$ thermocouples that were attached to the following points: a) the center back of each face of the hollow CCR (total of 3); b) the cylindrical case for the CCR (total of 1); c) the edge of the filter (total of 1); d) the 3 zones of the chamber (front, center, back) (total of 3); e) the window of the chamber (total of 1). 
2) The solar simulator was an incandescent bulb powered by a DC supply to a power level of 76~W.  This was placed at a distance of 75~mm from the cylinder containing the corner cube, giving a light intensity of 1075~W/m$^2$ approximately equal to sunlight at the lunar surface. Although the color temperature of this solar simulator does not match the color temperature of the Sun, the idea here was to shine a radiant heat source of approximately the same intensity as the Sun on the ``instrument'' (both, case and CCR), so as to simulate the temperature increase due to sunlight.
3) The vacuum in the chamber was in the range of $5\times 10^{-6}$ to $1\times 10^{-4}$ Torr.
4) The temperatures of the internal surfaces of the chamber (as seen by the experiment) were automatically controlled to achieve the desired temperatures of the case of the CCR.  These temperatures ended up being a few degrees above or below the temperature of the case as needed.  The internal surfaces of the chamber are painted black. Placing the CCR in a case (both in this experiment and on the Moon) was done to isolate the CCR from the environment, so that its temperature is uniform regardless of the emissivity or other characteristics of the environment.

During the test, the system's characteristics were read by a ZYGO interferometer\footnote{See details on ZYGO interferometer at \tt http://www.zygo.com/} to validate the design and evaluate the effect of thermal distortions on the diffraction pattern.  The measured test data on this smaller prototype was correlated with the output of the structural-thermal-optical-performance modeling code, thereby confirming the design assumptions and raising the overall technology readiness level of the instrument.

\section{Towards the Development of an advanced lunar retro-reflector}
\label{sec:development} 

Assembly of a hollow CCR requires attaching three precisely aligned flat mirrors.  The attachment technique needs to minimize stresses that would distort the optical figure of the CCR over the temperature range of the lunar day/night cycle, covering the range $\sim$90--390~K. The design requirements for a large hollow CCR for LLR are challenging; the dihedral angles of a CCR must be stable under a variable thermal environment. Variations of the dihedral angles directly affect the diffraction pattern of a CCR and therefore its signal strength.  Adhesives, optical contact, and mechanical clamping are known to introduce stresses from temperature variation, potentially creating a major technical problem.

One approach relies on the sodium hydroxyl-catalysis bonding technique that uses sodium hydroxide as an agent to promote bonding between the plates of the CCR.  This technique was developed at Stanford University for the GP-B mission for bonding components of its instrument assembly.  The catalytic bonding technique allowed critical arcsec accuracy alignments of optical telescope and gyroscope reference pieces at room temperature and the ability to maintain alignment when cooled to the operating temperature of 1.7~K. This performance was successfully demonstrated in flight. Recently the performance of hydroxyl-catalysis bonding technique was successfully tested to 500 K without failure. This temperature range/alignment tolerance is comparable with the stability accuracy, and thermal range needed for the CCR assembly.

The large CCR requires a specific optical pattern to account for the velocity of the terrestrial tracking station relative to the Moon.  The desired pattern can be achieved with a CCR assembly with a slight difference in one dihedral angle (0.8 arcsec) from a perfect CCR.  Achieving such tolerances, with one angle offset from 90$^\circ$, is a challenge for current CCR manufacturing capability. Furthermore, it takes about 2 minutes for the sodium hydroxyl-catalysis bonding to set. (Potassium hydroxide is another candidate bonding agent that may increase the alignment time. This option is currently being investigated.) Therefore, one needs to be able to assemble the CCR to the specified tolerances, verify the optical quality of the assembly (via a full-aperture ZYGO interferometer, for example), and correct any identified misalignments all during a short time. This introduces interesting challenges. 

\begin{wrapfigure}{R}{0.27\textwidth}
 \vspace{-10pt}
  \begin{center}
    \includegraphics[width=0.22\textwidth]{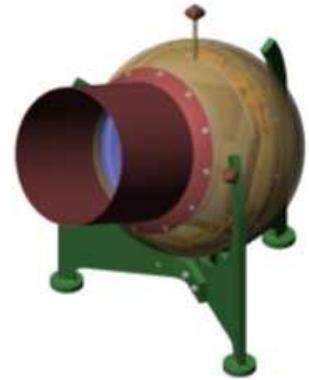}
  \end{center}
  \vspace{-10pt}
  \caption{LLR reflector housing with the MLI removed
shows two lift points for deployment at any targeted lander site on the Moon. Also visible: the CCR, narrow-pass-band solar filter with phase-corrector plate, solar shade and the deployment platform.}
\label{fig:16-CCR-Lunette}
  \vspace{-6pt}
\end{wrapfigure}

To alleviate this difficulty we will use an optical phase correction plate to modify the pattern of the assembled CCR, measured after manufacture, to correct it to the required pattern. The phase correction plate will be a flat plate perpendicular to the direction of the incoming/outgoing laser signal, polished across its surface to a variable thickness to produce small optical refractive delays as a function of position of the plate. Phase plates are often used in optical telescopes, though they do not appear to have been applied to a CCR assembly. 

To limit thermal distortions due to incident sunlight, both sides of the phase correction plate will be optically coated with sunlight rejection coatings to allow passage of 532 nm laser signals. The outer (sunward) surface of the corrector plate will be coated to reject light in the range 300--900 nm except for a narrow band 30~nm wide around the 532 nm wavelength. This will reject most of the heat from the sunlight. The phase plate substrate will attenuate sunlight at wavelengths shorter than 300~nm, resulting in some thermal leakage inside the container which is successfully mitigated by the overall thermal design of the instrument. The inside of the corrector plate will be coated to reflect light with wavelength longer than 900 nm. Filters of this type have been fabricated for similar purposes and tested previously, though for a different laser wavelength (1064 nm). 

Finally, to further minimize thermal gradients across the corner cube we will use MLI inside the container, thereby, minimizing possible optical distortions.  Detailed numerical simulations have been performed to show that a corner cube assembled with these technical innovations should, when suitably mounted, survive launch and give the required performance over the lunar cycle.  The analysis shows that the instrument will give reflected signal strength stronger than the fresh Apollo 11 and 14 arrays, while having low mass suitable for robotic deployment. 

\section{Deployment of the new instrument on the Moon}

The current design is intended for use on an unmanned lunar lander equipped with a robotic arm. As far as the deployment of the  new instrument on the Moon is concerned, it must be deployed on the lunar surface and away from a lander; it must point at the mean Earth to within 2 degrees and have the axis of the reflector optical axis parallel with the Earth's equator. Figure~\ref{fig:16-CCR-Lunette} shows the design that relies on the availability of a robotic arm on the lander.  Also, the clock angle (about the optical axis) must be set so as to align the two-lobed return pattern with the direction of velocity aberration.   The exact nature of the lunar surface (slope and roughness) will not be known prior to landing.  However, the elevation and clock angles relative to the local gravitational vertical will be known.

We have created an instrument design that takes advantage of this fact to set the elevation and clock angles, while allowing for variable lunar surface conditions. A spherical enclosure has been designed which provides lifting points for a lander robotic arm to pick it up and deploy it on the surface (Figure~\ref{fig:16-CCR-Lunette}).  The center of mass of the instrument is designed to coincide with the center of this sphere, thus the orientation of the optical axis will be in neutral equilibrium.  The alignment of the optical axis requires a different set of angles for each landing site, which is accommodated by having multiple lifting points for each enclosure.

After landing, the robotic arm will lift the enclosure using the lifting point specific to the actual landing location.  As the instrument is lowered into place, a separate tripod base (loosely attached to the spherical enclosure) will come to rest on the lunar surface first.  The CCR enclosure will rotate relative to the tripod base to the desired angle when lowered to the surface.  The robotic arm must deploy the enclosure to the correct azimuth before lowering it to the surface. Such a deployment is within the capabilities of several known future landers, such as those expected in Lunette \cite{Lunette:2010}, Luna-Globe\footnote{\tt http://www.russianspaceweb.com/luna\_glob.html} missions. 

The objective of any future installation on the Moon must be both to widen the array distribution on the lunar surface and to enable many SLR stations to achieve mm-level LLR ranging - a factor of 20 gain compared to present state. Increased sensitivity would allow a search for new effects due to the lunar fluid core free precession, inner core influences and stimulation of the free rotation modes. In addition to improved accuracy on the foregoing results, future possibilities include detection of an inner solid core interior to the fluid core.  Advances in gravitational physics are also expected. Also, the small number of current LLR stations could be expanded if the return signal was stronger. Therefore, we emphasize that the development and deployment of new CCR instruments on the Moon is well justified and, because of the aging of the CCR arrays currently on the Moon,  it must be realized in the near future.

Once deployed on the Moon, the new CCR instrument will be fresh and bright.  Since the laser pulse will not be spread out by the reflection, it will be possible to construct 1 mm accuracy normal points from fewer photons in a shorter time.  Thus, one can expect annual data rates to be at least comparable to current rates from the Apollo 15 array, making it a very attractive option for advancing many areas of the LLR-enabled science investigations \cite{Williams-etal-2009}.

\section{Conclusions and next steps}
\label{sec:concl}

We have discussed the development of a new LLR instrument based on a single large hollow CCR for which detailed design, analysis, and testing have been performed at JPL.  The new CCR (i) will allow for improving lunar range accuracy by more than an order of magnitude, (ii) will have large enough signal to be easily detected by many ground stations, and (iii) will have low enough mass to be deployed by a robotic arm. The proposed design for a new CCR instrument solves several issues (beam pattern, thermal distortion, and velocity aberration) that have discouraged the use of single CCRs previously.

When designing a new retro-reflector system for LLR, one must consider historical data which show that the strength of the signal returned by a lunar retro-reflector strongly affects the number of ranges acquired. Over four decades, the factor of 3 for the (Apollo 15)/(Apollo 11 or 14) signal strength ratio has affected the rate at which ranges are collected. Historical data indicate that the Lunokhod 2 signal has grown weaker with time. The case is less obvious for the Apollo arrays. The annual number of observations has not dramatically reduced, though this might be partly a result of improving technology. There is a lot that is not understood about levitated dust, but if levitated dust is fated to degrade any new retro-reflector with time then a strong initial signal is called for. We recommend the 170 mm design for future single-aperture LLR retro-reflector instruments. 

A strong reflected signal guarantees that scientifically useful ranges will be measured. A large single corner cube design can match or exceed the initial signal from the small Apollo arrays and may match Apollo 15.  Good thermal control and manipulation of the diffraction pattern improve performance. The design presented is capable of reaching the 1-mm Earth-Moon few-photon range precision (compared to a few cm currently) needed to initiate advanced LLR operations.  The instrument is expected to perform under the harsh environment on the lunar surface (e.g., dust, significant thermal variations), will have low mass, and will allow for robotic deployment.

Our design protects the CCR within an insulated case with a solar filter at the optical aperture.  This narrow-pass optical filter passes the laser ranging wavelength of 532 nm while minimizing the amount of sunlight incident on the reflective surfaces.  This leads to careful control of the temperature gradients within the CCR regardless of the lunar surface temperature and the position of the Sun in the lunar sky.

The design presented here is based on a carefully selected set of dihedral angles between faces of the corner cube to obtain a beam pattern optimized for use with lunar laser ranging. Because achieving the precise angles needed challenges current CCR manufacturing techniques, our design also allows for the addition of a phase-correction plate.  After the CCR is manufactured, its beam pattern is measured, and the plate is custom polished to correct any imperfections. We have created detailed thermal, mechanical, and optical numerical models to show that acceptable optical performance for a retro-reflector instrument with a coated, hollow CCR can be achieved for both lunar day and night operations using the approach described above.

Deployment of the new instrument on the Moon will continue the program of the LLR-enabled science investigations.

\begin{acknowledgments}
We thank Leon Alkalai, W. Bruce Banerdt, Hamid Hemmati, Michael Shao, and Michael Werner of JPL for their interest, support and encouragement during the work. We also thank David Arnold, Douglas Currie, and Thomas W. Murphy Jr. for helpful conversations. The work described in this report was performed at the Jet Propulsion Laboratory, California Institute of Technology, under a contract with the National Aeronautics and Space Administration. 
\end{acknowledgments}

\bibliography{ccr-paper}

\end{document}